\documentclass{aa}  

\usepackage{graphicx}
\usepackage{txfonts}
\usepackage{natbib,twoopt}
\usepackage{lscape}
\usepackage[colorlinks,citecolor=blue,urlcolor=blue]{hyperref} 
\bibpunct{(}{)}{;}{a}{}{,}             
\makeatletter
  \newcommandtwoopt{\citeads}[3][][]{\href{http://adsabs.harvard.edu/abs/#3}%
    {\def\hyper@linkstart##1##2{}%
     \let\hyper@linkend\@empty\citealp[#1][#2]{#3}}}
  \newcommandtwoopt{\citepads}[3][][]{\href{http://adsabs.harvard.edu/abs/#3}%
    {\def\hyper@linkstart##1##2{}%
     \let\hyper@linkend\@empty\citep[#1][#2]{#3}}}
  \newcommandtwoopt{\citetads}[3][][]{\href{http://adsabs.harvard.edu/abs/#3}%
    {\def\hyper@linkstart##1##2{}%
     \let\hyper@linkend\@empty\citet[#1][#2]{#3}}}
  \newcommandtwoopt{\citeyearads}[3][][]%
    {\href{http://adsabs.harvard.edu/abs/#3}
    {\def\hyper@linkstart##1##2{}%
     \let\hyper@linkend\@empty\citeyear[#1][#2]{#3}}}
\makeatother
\graphicspath{{./}{figures/}}
 \def\mso{\,\mathrm{M}_\odot}
 
 \def\lso{\,{\rm L}_\odot}
 \def\zso{\,{\rm Z}_\odot}

 \def\simle{\mathrel{\hbox{\rlap{\hbox{\lower4pt\hbox{$\sim$}}}\hbox{$<$}}}}
 \def\simgr{\mathrel{\hbox{\rlap{\hbox{\lower4pt\hbox{$\sim$}}}\hbox{$>$}}}}

\usepackage[colorlinks,citecolor=blue,urlcolor=blue]{hyperref}
\graphicspath{{./}{figures/}}
\usepackage{xcolor}
\definecolor{amaranth}{rgb}{0.9, 0.17, 0.31}

\usepackage{xcolor}
\definecolor{ochre}{rgb}{0.8, 0.47, 0.13}

\usepackage[normalem]{ulem}

\begin{document}

   \title{Stripped-Envelope Stars in Different Metallicity Environments}
   \subtitle{I. Evolutionary Phases, Classification and Populations}
   \titlerunning{Stripped-Envelope Stars in Different Metallicity Environments: Evolution, Classification and Populations}
   \authorrunning{Aguilera-Dena et al.}
   \author{David R. Aguilera-Dena\inst{1,2,3}
          \and
          Norbert Langer\inst{2,3}
          \and
          John Antoniadis\inst{1,3,2}
          \and
          Daniel Pauli\inst{4}
          \and
          Luc Dessart\inst{5}
          \and
          Alejandro Vigna-G\'omez\inst{6,7}
          \and
          G\"otz Gr\"afener\inst{2}
          \and
          Sung-Chul Yoon\inst{8}
          }

   \institute{Institute of Astrophysics, FORTH, Dept. of Physics, University of Crete, Voutes, University Campus, GR-71003 Heraklion, Greece\\
              \email{davidrad@ia.forth.gr}
         \and
            Argelander-Institut f\"ur Astronomie, Universit\"at Bonn, Auf dem H\"ugel 71, 53121 Bonn, Germany
         \and
             Max-Planck-Institut f\"ur Radioastronomie, Auf dem H\"ugel 69, 53121 Bonn, Germany
        \and
        Institut f\"{u}r Physik und Astronomie, Universit\"{a}t Potsdam, Karl-Liebknecht-Str. 24/25, 14476 Potsdam, Germany
         \and
             Institut d'Astrophysique de Paris, CNRS-Sorbonne Universit\'e, 98 bis boulevard Arago, F-75014 Paris, France
         \and
             Niels Bohr International Academy, The Niels Bohr Institute, Blegdamsvej 17, DK-2100 Copenhagen, Denmark
         \and
             DARK, Niels Bohr Institute, University of Copenhagen,
             Jagtvej 128, DK-2200, Copenhagen, Denmark
         \and
             Department of Physics and Astronomy, Seoul National University, Gwanak-gu, Seoul, 151-742, Republic of Korea
             }

  \abstract{
Massive stars that become stripped of their hydrogen envelope through binary interaction or winds can be observed either as Wolf-Rayet stars, if they have optically thick winds, or as transparent-wind stripped-envelope stars. We approximate their evolution through evolutionary models of single helium stars, and compute detailed model grids in the initial mass range 1.5 to 70 M$_{\odot}$ for metallicities between 0.01 and 0.04, from core helium ignition until core collapse. Throughout their lifetime, some stellar models expose the ashes of helium burning. We propose that models that have nitrogen-rich envelopes are candidate WN stars, while models with a carbon-rich surface are candidate WC stars during core helium burning, and WO stars afterwards. We measure metallicity dependence of the total lifetime of our models and the duration of their evolutionary phases. We propose an analytic estimate of the wind optical depth to distinguish models of Wolf-Rayet stars from transparent-wind stripped-envelope stars, and find that the luminosity ranges at which WN, WC and WO type stars can exist is a strong function of metallicity. We find that all carbon-rich models produced in our grids have optically thick winds and match the luminosity distribution of observed populations. We construct population models and predict the numbers of transparent-wind stripped-envelope stars and Wolf-Rayet stars, and derive their number ratios at different metallicities. We find that as metallicity increases, the number of transparent-wind stripped-envelope stars decreases and the number of Wolf-Rayet stars increases. At high metallicities WC and WO type stars become more common. We apply our population models to nearby galaxies, and find that populations are more sensitive to the transition luminosity between Wolf-Rayet stars and transparent-wind helium stars than to the metallicity dependent mass loss rates.
}


\maketitle
\defcitealias{aguilera-dena2021}{Paper II}
 
\section{Introduction} \label{sec:intro}

Many open questions in contemporary astrophysics are intertwined with the evolution and fate of massive stars. They are key in shaping their environments through the many feedback processes they experience, and they are the main source of chemical enrichment throughout cosmic history. The first generations of massive stars are believed to have contributed significantly to the re-ionization of the Universe \citep[e.g.][]{2020A&A...634A.134G}. The populations of massive stars in the more chemically enriched local Universe have a complex morphology \citep[e.g.][]{2011A&A...530A.108E,2021MNRAS.504.2968P}, and they are responsible for the formation of many astrophysical transients we observe \citep[e.g.][]{2019NatAs...3..717M}.

At the end of their evolution, massive stars can explode as core-collapse supernovae (SNe), leaving behind a neutron star (NS) or black hole (BH) remnant. Massive stars that do not produce successful explosions collapse, leaving behind BHs. Very massive stars can undergo pair instability, and above a certain mass they explode without leaving a remnant behind \citep{2012ARA&A..50..107L}. They input a large amount of mechanical and radiative energy in their environments, both through stellar winds on relatively long timescales, and through SNe in shorter timescales. As successive generations of stars are formed, their initial composition becomes more enriched in heavy elements produced by their predecessors. Understanding how their composition influences their evolution, energy output and final fate is therefore important for understanding galaxy evolution, as well as the properties of stellar populations, astrophysical transients such as SNe \citep[e.g.][]{2019NatAs...3..717M}, and compact object mergers \citep[e.g.][]{2018MNRAS.481.4009V} throughout the history of the Universe. 

Most massive stars in the Universe are thought to be in stellar binaries \citep[e.g.][]{2011IAUS..272..474S,2012Sci...337..444S} or multiples \citep[e.g.][]{2014ApJS..215...15S,2017ApJS..230...15M} that will interact at some point during their lifetime. Our understanding of the evolution of massive stars, and the role that multiplicity plays in their evolution in different environments, has taken grand leaps in the past decades \citep[e.g.][]{1992ApJ...391..246P,2010ApJ...725..940Y,2012ARA&A..50..107L,2021A&A...645A...5S,2021A&A...656A..58L}. However, many gaps remain in our understanding of the effect of metallicity and multiplicity on their evolution and fate.

Simulations of interacting binary systems indicate that at least one component star is likely to lose most or all of its hydrogen-rich envelope \citep[e.g.][]{1992ApJ...391..246P,2001A&A...369..939W,2017ApJ...840...10Y,2019MNRAS.486.4451G,2020A&A...637A...6L,2020A&A...638A..55K,2021arXiv210714526V}. This results in the formation of stripped-envelope stars, characterised by a surface that is hydrogen deficient and helium enriched.

Envelope stripping can occur due to mass transfer, e.g. during the main sequence (MS) evolution (Case A mass transfer; cf., \citealt{2021arXiv211103329S}), or following the depletion of hydrogen or helium in the core \citep[Case B and C mass transfer, respectively, see][for a review]{2006csxs.book..623T}. Envelope ejection can also occur on dynamical timescales, when a binary system undergoes a common envelope (CE) event that is often accompanied by a significant tightening of the binary orbit \citep{2001A&A...369..170T,2011ApJ...730...76I,2016A&A...596A..58K}. Furthermore, the most massive and luminous stars may lose their hydrogen envelopes without interacting with a companion star, either through winds during the MS evolution \citep{1975MSRSL...9..193C}, winds during the red supergiant phase \citep[][]{1981A&A...102..401M}, or massive explosive outbursts, when their outer layers are near the Eddington limit \citep[e.g.][]{1994A&A...290..819L}.

Several formation mechanisms of stripped-envelope stars may contribute to their numbers in the Universe, but their efficiency is uncertain \citep{2020A&A...634A..79S}. Regardless of the formation mechanism, stripped-envelope stars will have different surface properties and evolve differently than their non-stripped counterparts \citep{2019ApJ...878...49W,2021A&A...645A...5S,2021A&A...656A..58L}.

Some stripped-envelope stars will become classical Wolf-Rayet (WR) stars \citep{2014A&A...564A..30G}; a class of massive stars that is distinguished by spectra with strong emission lines, product of their high mass loss rates that lead to an optically thick wind above the stellar photosphere, and that show little or no hydrogen \citep{2007ARA&A..45..177C}. WR stars are observed in several sub-types that are indicative of their surface composition, temperature and evolutionary stage. The broadest classification divides them into WN, WC and WO type. WN type stars are characterised by the presence of nitrogen in their spectra, while WC stars have strong carbon and oxygen lines. Stars with these spectral types are believed to be undergoing helium burning in their core. WO stars have spectra dominated by oxygen lines, and are believed to have helium-depleted cores, placing them closer to the end of their evolution \citep{2015A&A...581A.110T}. The mass loss rates of WR stars have been found to be several orders of magnitude higher than those of main sequence stars, and they are strongly metallicity dependent \citep[e.g.][]{2005A&A...432..633G,2008A&A...482..945G,2005A&A...442..587V,2006A&A...457.1015H,2014A&A...565A..27H,2015A&A...581A.110T}.

The luminosities of WR stars have been found to have a lower limit, below which no WR stars are observed. This lower luminosity has been observed to be metallicity dependent. Understanding the formation channel of stars above and below this limit is not trivial, as they may originate both from single, wind-stripped stars, as well as from stars that have been stripped via an interaction with a companion \citep{2020A&A...634A..79S}. Theoretically, optically thick WR winds are expected to form above a given minimum luminosity, which depends on their metallicity \citep{2017A&A...608A..34G,2020MNRAS.499..873S}. Below this limit, the mass loss rates of stripped-envelope stars are expected to drop substantially \citep{2017A&A...607L...8V}, leading to a strong reduction or disappearance of their WR line emission. While this drop in wind mass loss rate still needs to be confirmed observationally, a similar effect is expected for continuous mass loss relations, where WR winds gradually become optically thin for lower masses. \cite{2018A&A...615A..78G} have shown that, due to this effect, low mass stripped-envelope stars emit most of their radiation in the UV, and are therefore difficult to detect. The study of WR stars through direct observation has been a very active field since their discovery \citep{2019Galax...7...74N}, but the study of stripped-envelope stars below the WR luminosity limit is a field that is expected to grow with upcoming observational campaigns. However, detailed studies are limited to nearby galaxies where systems can be observed individually.

In a recent study, \cite{2017MNRAS.470.3970Y} found that an increase in the mass-loss rate observed in WC stars could account for the formation of the faintest WC and WO type stars in the LMC and in our Galaxy. WR winds are known to have a strong metallicity dependence \citep[][]{2014A&A...565A..27H,2015A&A...581A.110T}, which strongly impacts their populations in different galaxies.

With this in mind, we study the evolution of helium stars as a proxy for stripped-envelope stars, including wind mass loss, in a similar fashion to \cite{2017MNRAS.470.3970Y} and \cite{2019ApJ...878...49W}. We extend the considered range of initial conditions to cover different metallicity environments, focusing on high metallicities, which are not often addressed in the literature, and can explain several of the observed population properties of WR stars and Type I core-collapse SN
 \citep[discussed in][hereafter \citetalias{aguilera-dena2021}]{aguilera-dena2021}

We pay particular attention to the empirical, metallicity-dependent lower luminosity limit of WR stars \citep{2020A&A...634A..79S}, apparent when comparing the WR populations in the Small and Large Magellanic Clouds \citep[e.g.][]{2003PASP..115.1265M,2014ApJ...788...83M,2018ApJ...863..181N} and the Galaxy \citep[e.g.][]{2019A&A...625A..57H}. This limit is likely related to the predicted drop of the mass loss rate for lower-luminosity helium stars \citep{2017A&A...608A..34G,2017A&A...607L...8V,2020MNRAS.499..873S}. 

We have subdivided our paper as follows: In Sect. \ref{sec:methods} we describe the physical and numerical treatment we use in our simulations, as well as a method we propose to find the minimum luminosity of WN and WC stars as a function of metallicity. In Sect. \ref{sec:results} we present how metallicity affects the time-dependent properties of our stripped-envelope star models. In Sect. \ref{sec:tau_analysis} we contrast the results of our numerical models to the inferred minimum luminosities of WR stars. In Sect. \ref{sec:populations}, we show the implications of our results for the morphology of WR populations. We finalise our paper with a discussion in Sect. \ref{sec:discussion} and conclusions in Sect. \ref{sec:conclusions}. In a companion paper \citepalias{aguilera-dena2021} we discuss the properties of our stripped-envelope stellar models at core-collapse, and present an analysis of the transients and the remnants we expect from them. Additionally, some of the ramifications for compact-object mergers are also discussed in \cite{2022A&A...657L...6A}.

\section{Method}\label{sec:methods}

\begin{table*}
\caption{Notation definitions.}\label{tab:notation}
\centering
\begin{tabular}{ll}
\hline \hline 
Symbol & Definition \\
\hline 
$\mathrm{L}^\mathrm{tau}_\mathrm{min,WN}$ & Minimum luminosity above which a stellar model is classified as a WN star, according to Sect. \ref{sec:tau_method}  \\
$\mathrm{L}^\mathrm{tau}_\mathrm{min,WC}$ &  Minimum luminosity above which a stellar model is classified as a WC star, according to Sect. \ref{sec:tau_method}\\
$\mathrm{L}^\mathrm{evo}_\mathrm{min,WC}$ & Minimum luminosity where a WC is produced by winds in our evolutionary calculations \\
$\mathrm{L}^\mathrm{evo}_\mathrm{min,WO}$ & Minimum luminosity where a WO is produced by winds in our evolutionary calculations \\
$\mathrm{t}_\mathrm{total}$ & Total lifetime of a helium star from core helium ignition to core collapse \\
$\mathrm{t}_\mathrm{He-N}$ & Lifetime of a helium star spent with an outer helium-nitrogen shell \\
$\mathrm{t}_\mathrm{He-C}$ & Lifetime of a helium star spent with an outer helium-carbon shell\\
$\mathrm{t}_\mathrm{He-O}$ & Lifetime of a helium star spent with oxygen surface abundance larger than 0.05 after core helium depletion\\
\hline
\end{tabular}
\end{table*}

In this paper, we take several approaches to study the effect of metallicity on the evolution of stripped-envelope stars, all of which are detailed below. In Sect. \ref{sec:models} we describe the method employed to construct a grid of stellar evolution models of helium stars, meant to approximate the evolution of stripped-envelope stars. In Sect. \ref{sec:tau_method} we present a semi-analytic method we devised to estimate the wind optical depth of stripped-envelope stars. We then use this criterion (in Sect. \ref{sec:tau_analysis}) to classify our models as either WRs (with optically thick winds) or transparent-wind helium stars. Finally, we combine the models and the semi-analytical method to produce synthetic populations of stripped-envelope stars in different metallicity environments, through the method detailed in Sect. \ref{sec:methods_popsynth}.

To simplify the presentation of our results, we introduce the notation used in this paper for various useful quantities in Table\,\ref{tab:notation}. 

\subsection{Stellar evolution models}\label{sec:models}
We computed grids of evolutionary sequences of non-rotating helium stars using version 10398 of the Modules for Experiments in Stellar Astrophysics (\texttt{MESA}) code \citep{MESAI,MESAII,MESAIII,2018ApJS..234...34P}, from the beginning of core helium burning to the onset of core collapse (when possible), defined as the point where the infall velocity at the edge of the iron core reaches ${1000\,\mathrm{km\,s^{-1}}}$.

To test the effect of initial chemical composition in hydrogen depleted stripped-envelope stars, we performed calculations for 7 different metallicities ($Z_{\rm init}$), from 0.01 to 0.04 in steps of 0.005, scaled from solar abundances found by \cite{1996ASPC...99..117G}, and initial masses between 1.5 and 70 $\mso$ in steps of 0.5 $\mso$, which correspond to the helium core masses of stars with zero-age main sequence (ZAMS) masses between roughly 10 and 150 $\mso$ \citep[][see Eq. \ref{eq:zams_he}]{2019ApJ...878...49W}.

The initial models were generated from hydrogen-rich pre-main sequence models at the aforementioned metallicities. They were evolved imposing chemical homogeneity by means of an artificially imposed, very high mixing coefficient throughout the star, and without mass loss, until the end of hydrogen burning, but before the ignition of the $^{14}$N($\alpha$,$\gamma$)$^{18}$F($\beta^{+}$,$\nu$)$^{18}$O($\alpha$,$\gamma$)$^{22}$Ne reaction in the core; a reaction which takes place during a short timescale and at a lower temperature than that of helium burning, and depletes the core of nitrogen. This was done to ensure that nitrogen was not depleted in the stellar surface at the beginning of the calculation. The models were evolved further until they settled into thermal equilibrium from helium burning, relaxing the condition of homogeneity. This provides the abundances of CNO isotopes in the envelope that correspond to those of the core of a massive star after hydrogen burning (enhanced N, reduced C and O), and provides a helium content given by $\mathrm{Y}_\mathrm{init}=1-\mathrm{Z}_\mathrm{init}$.

Although binary evolution models typically retain residual hydrogen in their envelope after having been stripped due to Roche lobe overflow \citep{2017ApJ...840...10Y, 2019MNRAS.486.4451G,2020A&A...637A...6L}, we assume that the remaining amount of hydrogen is small, and will be removed in a short time scale by winds or case B mass transfer, and therefore does not significantly affect the evolution of helium stars (see Sect. \ref{sec:init_conds} for a discussion).

Convection was modelled using the standard mixing length theory \citep{1958ZA.....46..108B}, using the default MESA value of $\alpha_{MLT} = 2.0$, and adopting the Ledoux criterion for instability. We employed efficient semiconvection, using $\alpha_{SC} = 1.0$, following the results of \cite{2019A&A...625A.132S}. Convective overshooting was not included during any of the phases of evolution. We included thermohaline mixing as prescribed by \cite{2010A&A...521A...9C}. To improve the numerical stability of our calculations, we employed MESA's predictive mixing in helium burning shells \citep{MESAIII}. We calculated the energy generation rates and chemical composition changes using MESA's \texttt{approx21} nuclear network, from the initial model to core collapse.

Since helium stars often experience instabilities in their envelopes due to the proximity to the Eddington limit (e.g. \citealt{2015A&A...580A..20S}), our models were calculated using efficient energy transport through MESA's \texttt{mlt++} \citep{MESAII}, effectively increasing the efficiency of convection to transport energy in convective layers in the envelope. In some cases, in order to produce models that converge up to core collapse, we exclude radiative acceleration in the envelope by setting the radial velocity to 0 in layers with $T<10^8$ K during the late evolution, when timesteps become smaller than 0.1 yr. Additional numerical details are available online\footnote{\url{https://zenodo.org/deposit/5747933}}.

To compute the mass loss that our models experience during their evolution, we use the mass loss recipes suggested by \cite{2017MNRAS.470.3970Y}, which prescribe different mass loss rates for WN and WC type stars.
These recipes are based on the work of \cite{2014A&A...565A..27H} for WN stars, using the metallicity dependence obtained by \cite{2006A&A...457.1015H}; and \cite{2016ApJ...833..133T} for WC/WO stars. The mass loss rate recipe employed has the form
\begin{equation}\label{eq:windWN}
\dot{\text{M}}_{\text{WN}} = f_{\text{WR}} \left(\frac{\text{L}}{\lso}\right)^{1.18}  \left(\frac{\text{Z}_{\text{init}}}{\zso}\right)^{0.6} 10^{-11.32} \frac{{\text{M}_{\odot}}}{\text{yr}},
\end{equation}
for Y=1-Z$_{\text{init}}$, while for Y$<$0.9 it is given by
\begin{equation}\label{eq:windWC}
\dot{\text{M}}_{\text{WC}} = f_{\text{WR}} \left(\frac{\text{L}}{\lso}\right)^{0.85}  \left(\frac{\text{Z}_{\text{init}}}{\zso}\right)^{0.25} \text{Y}^{0.85} 10^{-9.2} \frac{{\text{M}_{\odot}}}{\text{yr}}.
\end{equation}
As suggested by \cite{2017MNRAS.470.3970Y}, we interpolate between them linearly in the regime between $\mathrm{Y}=1-\mathrm{Z}_{\text{init}}$ and $\mathrm{Y}<0.9$ using
\begin{equation}
\dot{\text{M}} = (1-x)\dot{\text{M}}_{\text{WN}} + x\dot{\text{M}}_{\text{WC}},
\end{equation}
with $x=(1-\text{Z}_{\text{init}}-\text{Y})/(1-\text{Z}_{\text{init}}-0.9)$. We take the recommended value of $f_{\text{WR}}$=1.58, corresponding to a clumping factor of $\mathrm{D}=4$, and scale these equations assuming $\zso=$0.02. These wind mass loss rates for WR stars were determined empirically, and are likely to be improved in the future. Some of the helium stars we model are outside of the observed luminosity range where WR stars have been observed. We are therefore effectively extrapolating WR mass loss rates to lower luminosities, which may have lower mass loss rates \citep{2017A&A...607L...8V}. We take this into account \textit{a posteriori} during the analysis of our results.

\subsection{Stellar wind optical depth}\label{sec:tau_method}

\cite{2020A&A...634A..79S} found the minimum luminosity of WR stars in the LMC, the SMC and the Galaxy, and propose an analytical expression for this quantity as a function of metallicity. They derive it by means of the so-called ``transformed radius'' $R_t$ \citep{1989A&A...210..236S}, which depends on the stellar radius, terminal wind velocity, mass loss rate and wind clumping factor. They assume that the mass loss rate depends on luminosity and metallicity with a functional form given by $\mathrm{\dot{M}}\propto \mathrm{L}^{\alpha} \mathrm{Z}^{\beta}$, and all other quantities are independent of either $\mathrm{L}$ or $\mathrm{Z}$. Finally, they find the minimum luminosity of WR stars by assuming that there is a metallicity-independent value of $R_t$ above which stripped-envelope stars will have WR spectra. They determine that this minimum luminosity is proportional to $\mathrm{Z}^{4\alpha / (3-4\beta)}$. Using values for $\alpha$ and $\beta$ derived by \cite{2017A&A...607L...8V} in the context of low mass helium stars, with luminosities below the minimum observed WN luminosity, they find a power law relation between the luminosity threshold and $\mathrm{Z}$ with an exponent is roughly equal to -1, which provides a reasonable fit to the observed values.

Here, we take a similar approach, but instead calculate the optical depth at the base of the winds of WR stars, $\tau (R)$, using the analytical approach of \cite{1989A&A...210...93L}. The minimum luminosity of WR stars should then be determined by the minimum luminosity at which WR winds are optically thick (i.e. have $\tau(R) \gtrsim 1 $). In this formalism, for a wind velocity law with $\beta=1$ \citep[Eq. 8 in][]{1989A&A...210...93L}, the optical depth at the base of the wind is given by
\begin{equation}\label{eq:tau1}
\tau (R) = \frac{\kappa |\dot{M}|}{4\pi R (v_\infty - v_0)} \ln \frac{v_\infty}{v_0},
\end{equation}
where $\kappa$ is the opacity, $R$ the stellar radius, and $v_\infty$ and $v_0$ are the terminal wind velocity and the velocity at the base of the wind, respectively.

We then combine this with the metallicity- and luminosity-dependent wind mass loss rates for both WN and WC type stars with Eqs.\,\ref{eq:windWN} and \ref{eq:windWC}, thereby extending the scope of \cite{2020A&A...634A..79S} to also include WC type stars. 

We assume that the terminal wind velocity is given by the escape velocity at the surface, multiplied by a factor $K$ \citep{2017A&A...608A..34G}; namely,
\begin{equation}
   v_{\infty} = K \times v_{esc} = K \sqrt{\frac{2GM}{R}(1-\Gamma}),
\end{equation}
where $\Gamma$ is the Eddington factor. Substituting these values in Eq.\,\ref{eq:tau1} yields:
\begin{eqnarray}\label{eq:tau2}
\begin{array}{cc}
    \tau (L,Z)= &\frac{\kappa |\dot{M}(L,Z)|}{4\pi R(L)}\left(K \sqrt{\frac{2GM(L)}{R(L)}(1-\frac{\kappa L}{4\pi Gc M(L)})}-v_0\right)^{-1} \\
    &\times \ln \left(\frac{1.3 \sqrt{\frac{2GM(L)}{R(L)}(1-\frac{\kappa L}{4\pi Gc M(L)})}}{v_0}\right).
\end{array}
\end{eqnarray}

We employ the mass-luminosity and radius-luminosity relations from \cite{1989A&A...210...93L}, assume that the opacity at the base of the wind is given by the electron scattering opacity, ${\kappa = 0.2  \,\text{cm}^2 \text{g}^{-1}}$, and that $v_0 = 20 \,\mathrm{km\,s^{-1}}$, representative of the sound speed in the envelopes of WR stars. After making these assumptions, the optical depth at the base of the wind of a WR star is only a function of luminosity and metallicity, and it is different for WN and WC type stars.

\begin{figure}
\resizebox{\hsize}{!}{\includegraphics{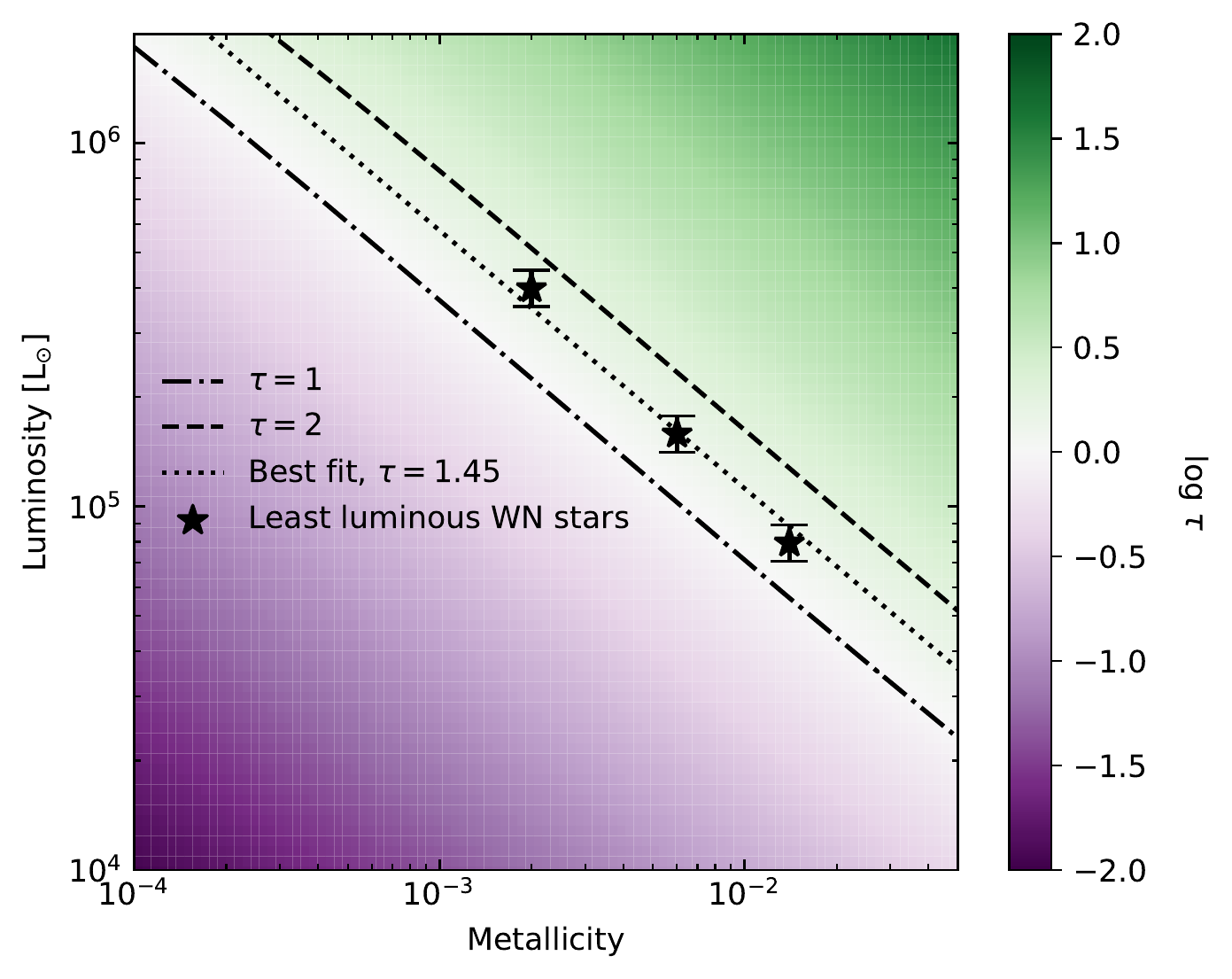}}
\caption{Optical depth at the base of the wind of a stripped-envelope star as a function of its metallicity and luminosity, obtained from Eq.\,\ref{eq:tau2}. Based on the optical depth derivation, mass-luminosity, and radius-luminosity relations of \cite{1989A&A...210...93L}. We employ the mass loss rate of WN stars, which depends on metallicity and luminosity as in Eq. \ref{eq:windWN}. Star shaped symbols represent the lowest luminosity observed for WN stars in the SMC, LMC and Milky Way, according to \cite{2020A&A...634A..79S}. Error bars with a width of 0.05 dex are included, as these values are rounded.
The lines represent the solutions of Eq. \ref{eq:tau2} for constant optical depth, including those that best fit the observed values of the observed minimum luminosities of WN stars.
} \label{fig:tau}
\end{figure}

The result of this calculation assuming the mass loss rate of WN stars (Eq. \ref{eq:windWN}), and using $K=1.3$ \citep{2011A&A...535A..56G} is presented in Fig. \ref{fig:tau}. The dashed-dotted line corresponds to the region where $\tau=1$. This line separates the optically-thick region (green), where we expect stripped-envelope stars to have WR spectra, and the optically thin region (purple). Since this model relies on several simplifying assumptions, and the uncertainties they introduce cannot be accounted for in detail, we find the values of the minimum luminosities of each type of WR by fitting a line of constant optical depth to the observed lower luminosity values of WR stars as a function of metallicity. For WN type stars, we calibrate our calculation to the observed values of the minimum luminosity of WN stars in the SMC, LMC and the Galaxy from \cite{2020A&A...634A..79S}.

We find that the observed minimum luminosities of WN stars lie at optical depths larger than 1, but of order unity, and conclude from this that our approach can be used to estimate the threshold WN luminosity as a function of metallicity. We find that the curve described by setting $\tau = 1.45$ provides the best fit to the data. This is equivalent to changing the value of any of the constants whose value we (somewhat arbitrarily) set in Eq. \ref{eq:tau2}, to make the curve with $\tau=1$ coincide with the data. We highlight that neither approach is preferable to the other, but we argue that by setting reasonable values for the unknown quantities and finding relative good agreement with our expectations justifies the validity of this method.

\begin{figure}
\centering
\resizebox{\hsize}{!}{\includegraphics{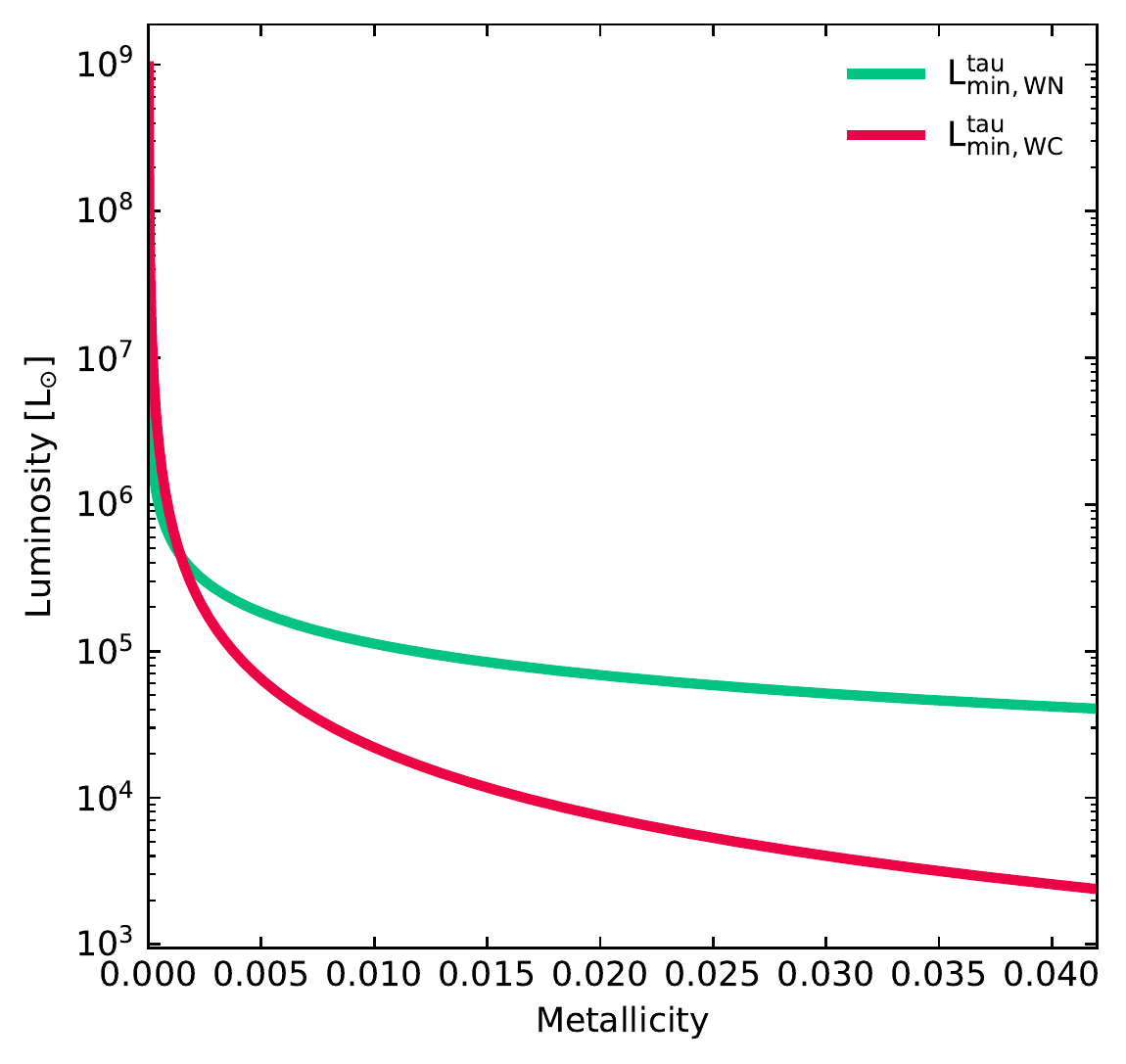}}
\caption{Minimum luminosity at which stripped-envelope stars can be observed as WN (green) and WC (red) type stars, as a function of metallicity, according to Eqs. \ref{eq:fit_wn} and \ref{eq:fit_wc}, respectively. 
}\label{fig:lums_both}
\end{figure}

We then find the minimum luminosity for WC stars as a function of metallicity by assuming they are described by a curve with the same optical depth, but with the mass loss rate given by Eq. $\ref{eq:windWC}$, and a value of $K = 1.6$ \citep{2011A&A...535A..56G}. The values of the minimum luminosities of WN and WC stars found using this method, labelled $\mathrm{L}^\mathrm{tau}_\mathrm{min,WN}$ and $\mathrm{L}^\mathrm{tau}_\mathrm{min,WC}$ respectively, are shown in Fig. \ref{fig:lums_both}. 

Figure \ref{fig:tau} shows that the curves of constant optical depth are well described by power laws between the luminosity and the metallicity. For WN type stars, we find that the best fit to the observations is given by
\begin{equation}\label{eq:fit_wn}
    \mathrm{L}^\mathrm{tau}_\mathrm{min,WN} = 6.85 \times 10^{4} \left(\frac{Z}{0.02}\right)^{-0.71} \lso,
\end{equation}
and for WC type stars, the curve with $\tau = 1.45$ is described by
\begin{equation}\label{eq:fit_wc}
    \mathrm{L}^\mathrm{tau}_\mathrm{min,WC} = 7.51 \times 10^{3} \left(\frac{Z}{0.02}\right)^{-1.55} \lso.
\end{equation}

We find that curves of constant optical depth provide a better fit to the observed minimum WN type star luminosities than a power law with slope of $-1$. Furthermore, by extending this analysis to WC stars, we are also able to compare the minimum luminosity of WC stars predicted by our optical depth model to the minimum luminosity at which WC stars are observed, as well as the minimum luminosity of the carbon-rich models that are produced in our grids.

The proposed method to find the minimum luminosities of WR stars has been employed by \cite{pauli}, who find a similar calibration, and extend it to classify H-rich WR stars, obtained in binary evolution models. Their binary models are made at LMC metallicity, and they find that the number and the luminosity distribution of hydrogen-rich WN stars, hydrogen-poor WN stars and WC stars are well reproduced by their models. They also predict core hydrogen burning stars that develop optically-thick winds. Models in their grid that expose the products of helium burning, are found to always be luminous enough to be classified as WC stars. Our method is in agreement with their findings as no carbon-rich stars with transparent winds are predicted by our models (see Sect. \ref{sec:tau_analysis}).

We note that this method for distinguishing WR stars from low mass stripped-envelope stars with transparent winds is agnostic to whether the mass loss rates change abruptly between one type and the other or not.

\subsection{Population synthesis models}\label{sec:methods_popsynth}

Using the models described in Sec. \ref{sec:models} and the semi-analytical method described in Sec. \ref{sec:tau_method}, we construct simple synthetic populations of stripped-envelope stars in different metallicities. We distinguish between models with surface optical depth larger or smaller than $\tau=1.45$, classifying the former as WR stars, and the latter as transparent-wind stripped-envelope stars. We classify a model as a WN star if it has $\tau > 1.45$ and shows nitrogen on the surface. If surface nitrogen is depleted and surface carbon enhanced, it is classified as WC. When a model has a surface oxygen abundance $X_\mathrm{O} > 0.05$ and the model has completed core helium burning, it is classified as a WO star \citep[][]{2015A&A...581A.110T}.

We use these model stellar populations to estimate how the number of WR stars of the different types we consider, and transparent-wind stripped-envelope stars, change as a function of metallicity. We also compute how the average number ratios of WC to WN stars, $\langle \text{N}_{\text{WC}}/\text{N}_{\text{WN}}\rangle$, WO to WN stars, $\langle \text{N}_{\text{WO}}/\text{N}_{\text{WN}}\rangle$, and WO to WC stars, $\langle \text{N}_{\text{WO}}/\text{N}_{\text{WC}}\rangle$ vary as a function of metallicity.

We construct the stellar populations assuming a constant star formation rate and an initial mass function (IMF) given by $\xi(\mathrm{M}_{\mathrm{ZAMS}}) \propto \mathrm{M}_{\mathrm{ZAMS}}^{-2.35}$ \citep{1955ApJ...121..161S}. For each metallicity, we draw 5000 stars with masses given by the IMF every 1000 yr, for a total of 60 Myr, totalling $3\times 10^{7}$ stars. We assume that the ZAMS mass of a star that produces a helium core of a given mass (which we later assume becomes a helium star) is not a strong function of metallicity, but only of the initial helium star mass. We employ the formula of \cite{2019ApJ...878...49W}, given by
\begin{eqnarray}\label{eq:zams_he}
\frac{\mathrm{M}_{\mathrm{He,ini}}}{\mso}  & \approx&
 \left\{
\begin{array}{cc}
0.0385 \ \mathrm{M}_{\mathrm{ZAMS}}^{1.603}, & \mathrm{if}  \ \mathrm{M}_{\mathrm{ZAMS}} < 30 \mso \\
0.5 \textrm{M}_{\mathrm{ZAMS}} - 5.87, & \textrm{if} \ \mathrm{M}_{\mathrm{ZAMS}} \geq 30 \mso . \\
\end{array}     \right.
\end{eqnarray}

We then assume that the ZAMS lifetime of these stars is given by their nuclear timescale, and that all stars are immediately stripped at the beginning of core helium burning, and evolve as single helium stars.  
We use the values of  $\mathrm{t}_\mathrm{He-N}$, $\mathrm{t}_\mathrm{He-C}$ and $\mathrm{t}_\mathrm{He-O}$ found from our models to calculate the number of each type of stripped-envelope star as a function of time. 

The maximum ZAMS mass in our populations is taken as 151.74 $\mso$, corresponding to the ZAMS mass of our most massive helium star model, with 70 $\mso$. Due to the slope of the IMF, these results are not sensitive to the exact value of the maximum mass as long as it is large. The minimum mass for the transparent-wind stripped-envelope stars that are tracked in our populations is of $3 \mso$, which corresponds to a ZAMS mass of approximately $15.14  \mso$. This is chosen to reflect the minimum mass of stripped-envelope stars that undergo core-collapse \citep{2019ApJ...878...49W,2022arXiv220100871C}.

\section{The impact of metallicity on the evolution of stripped-envelope stars}\label{sec:results}

In this Section we study the effect of metallicity on the evolution of stripped-envelope stars by analysing the grids of evolutionary calculations described in Section \ref{sec:models}. 

We have subdivided this Section as follows: In Section \ref{sec:evol} we review the evolution of helium star models of different masses and metallicities with a few characteristic examples. In Section \ref{sec:lifetimes} we characterise their lifetimes, and the duration of their different evolutionary phases. Finally, in Section \ref{sec:lums} we present the luminosity ranges at which carbon- and oxygen-rich stellar models are produced as a function of metallicity.

\begin{figure*}
\centering
\includegraphics[width=8cm]{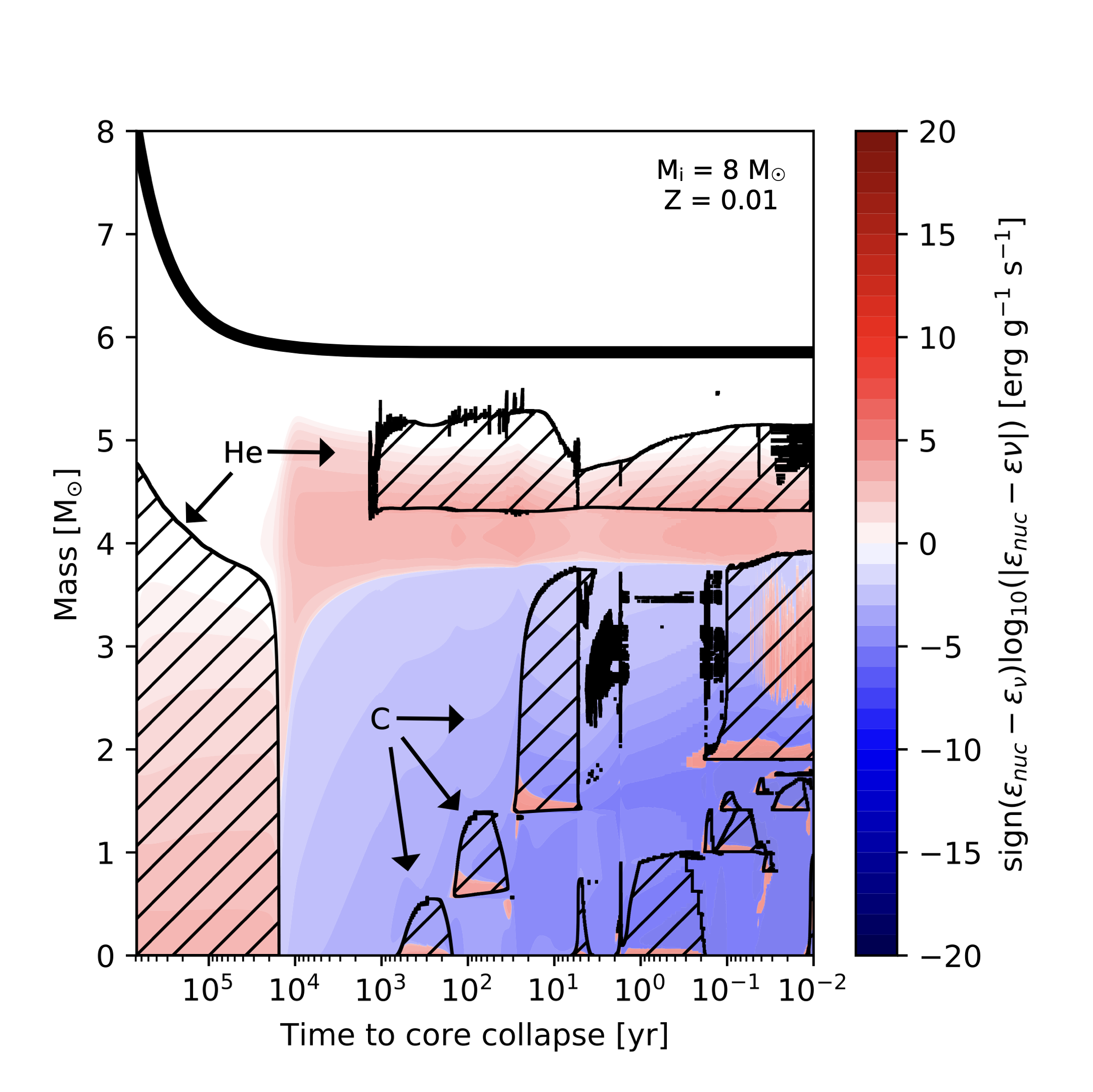}
\includegraphics[width=8cm]{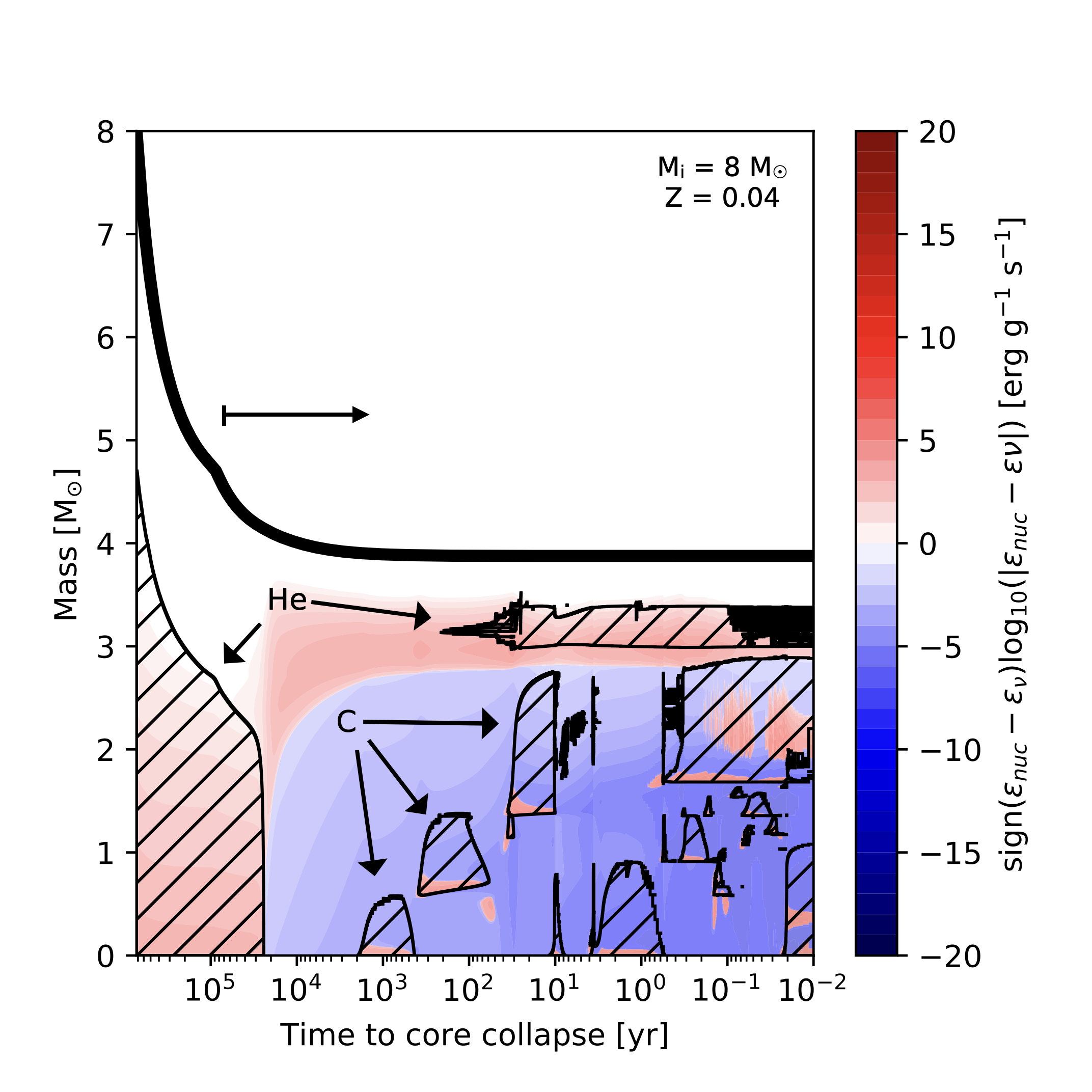}
\includegraphics[width=8cm]{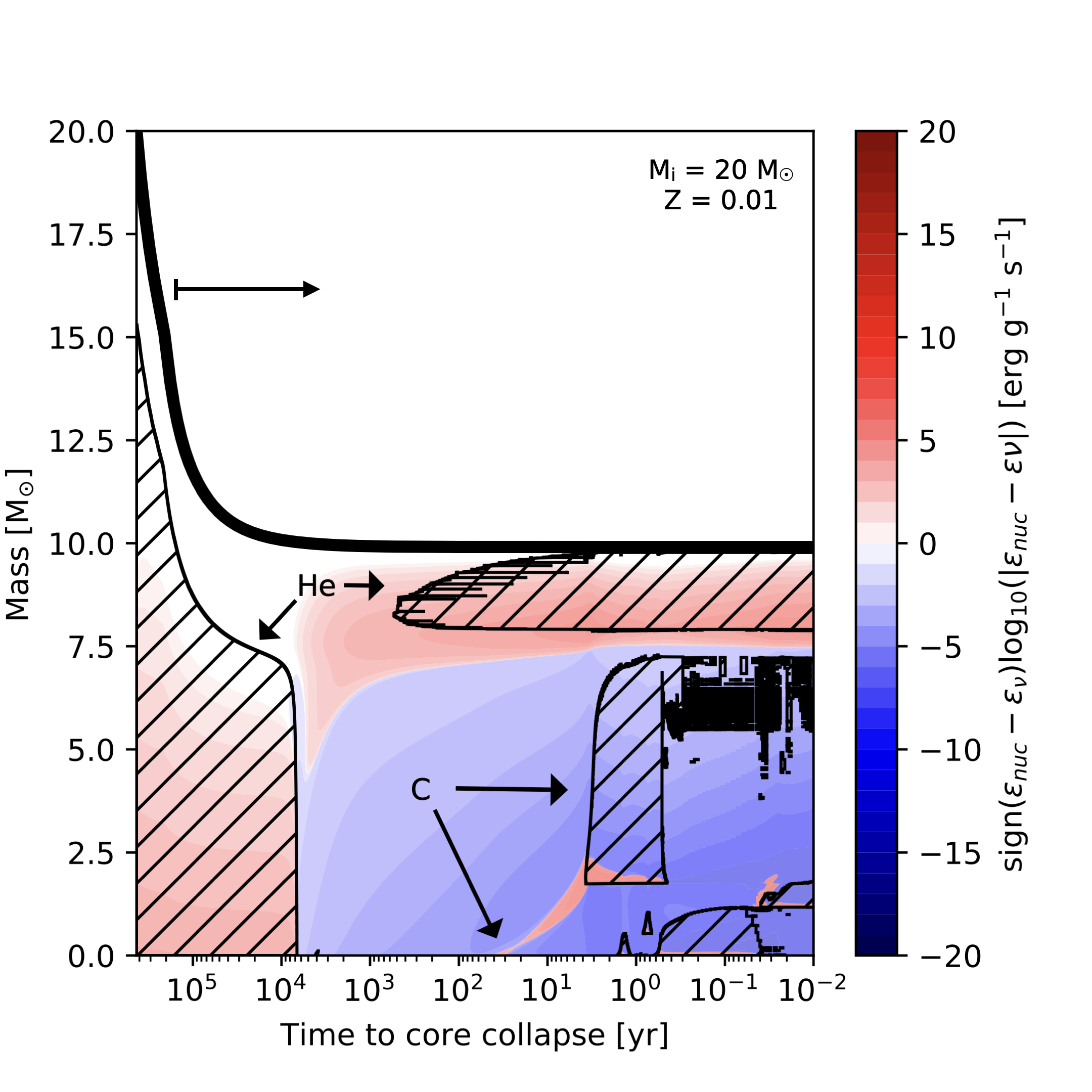}
\includegraphics[width=8cm]{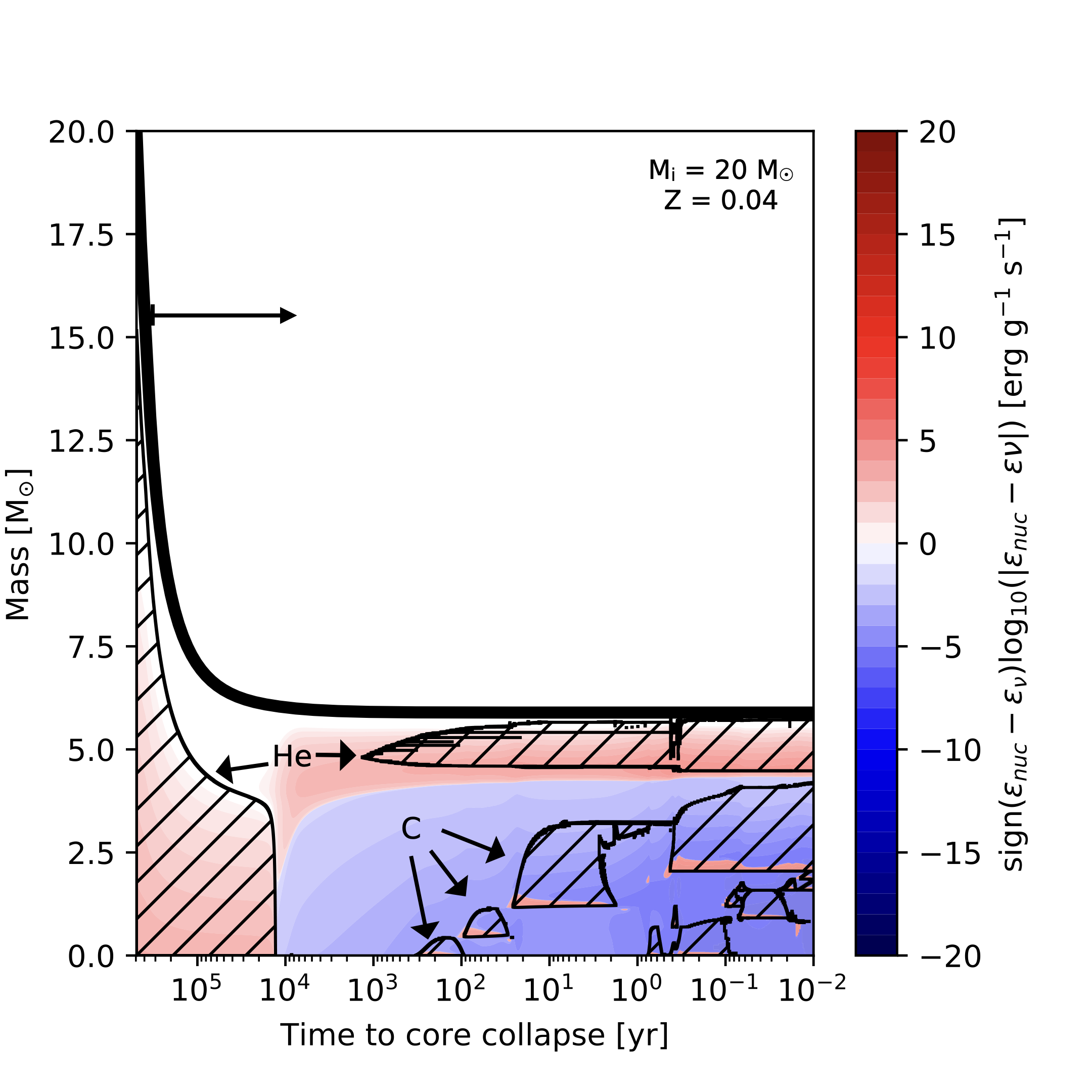}

\caption{Kippenhahn diagrams following the energy generation/loss rate and the structure of convective regions as a function of time remaining until core collapse,
from core helium burning until a few days before core collapse. Represented are the evolutionary calculations with initial helium star masses of 8 (top) and 20 (bottom) $\mso$, with metallicities of $Z=0.01$ (left) and $Z=0.04$ (right).
Colour denotes the intensity of the energy generation (red) and loss (blue) rate. Hatched regions denote convective regions. The helium and carbon burning regions are signalled with arrows. The point where the products of helium burning are exposed at the surface, and the model transitions from WN to WC mass loss rates is indicated with horizontal arrows.
\label{fig:kips}}
\end{figure*}

\subsection{Representative examples}\label{sec:evol}

Helium stars that are massive enough to burn helium in their core through the triple alpha process begin their core helium burning lifetime with a convective core and a predominantly radiative envelope. The latter is composed mostly of helium and nitrogen at this stage. Its carbon, nitrogen and oxygen abundances are set during hydrogen burning by CNO equilibrium. The core is mostly composed of helium, but is enriched with neon and depleted of nitrogen, since nitrogen burns at a lower temperature than helium. As helium burning progresses, the convective helium-burning core becomes increasingly richer in carbon, but also becomes enriched in oxygen through alpha captures on carbon. The total amount of oxygen produced from this reaction, however, remains uncertain because this reaction rate in particular is poorly constrained \citep[e.g.][]{2020ApJ...902L..36F}.

In our models, the transition period between the phases where a nitrogen-rich envelope is present, and the carbon-rich layers are exposed, is short. Nitrogen is thereafter absent from their surface, replaced by the products of helium burning. As mass loss continues to uncover deeper layers, the amount of carbon in the surface will decrease and the amount of oxygen will increase, since deeper layers of the stars spent a longer time exposed to the ashes of the $^{12}$C($\alpha$,$\gamma$)$^{16}$O reaction.

Depending on their mass and metallicity, helium stars can reach different outcomes at the end of their evolution. Possible outcomes have been characterised and studied in detail at solar metallicity \citep[][]{2019ApJ...878...49W,2020ApJ...890...51E,2022arXiv220100871C}. Helium stars that are massive enough to burn helium, but with initial masses below $\sim 1.9 \mso$ become CO or ONeMg white dwarfs (WDs). A small mass window may exist where helium stars go through more advanced nuclear burning processes that result in the formation of thermonuclear and electron-capture SNe \citep{2020A&A...635A..72A,2022arXiv220100871C}. Helium stars more massive than about $\sim 2.8 \mso$ form an iron core that will eventually collapse. Helium stars with final masses larger than $\sim 35 \mso$ may form either pulsational pair-instability SNe, leading to a core collapse and the formation of a BH, or a pair-instability SN, leading to an explosion that leaves no remnant behind \citep[e.g.][]{2017ApJ...836..244W}.

With the physical and numerical choices we have taken, mass loss is the process that most drastically affects the evolution of helium stars of different metallicities. The effect on envelope inflation is not captured by our treatment using MESA's \texttt{mlt++}.

Some models lose enough mass to expose the layers where nitrogen is depleted and carbon and oxygen are enhanced. At this point, they experience an increase in mass loss rate, corresponding to the WC stage. This creates a dichotomy in surface chemical composition of helium star models at core collapse, as well as a distinct distribution of final masses (\citealt{2017MNRAS.470.3970Y}; \citetalias{aguilera-dena2021}).

Since helium stars lose mass at a high rate, above a certain initial mass their convective cores tend to decrease in mass. Therefore, no semiconvective layers are formed in regions with chemical composition gradients. This, along with the fact that our models do not include rotational mixing, or mixing by tides, implies that stars that could be observable as WN/WC type stars, with surface layers rich in both carbon and nitrogen, are not produced in our grids \citep{1991A&A...248..531L,2021MNRAS.503.2726H}.

The evolution of helium star models of different mass and metallicity is illustrated by the Kippenhahn diagrams of four example evolutionary calculations in Fig. \ref{fig:kips}. The upper left panel shows our 8 $\mso$ model with a metallicity of 0.01 (corresponding to a star with ZAMS mass of about $28\mso$). It has a convective core that initially spans $\sim$4.8$\mso$, and which gradually decreases in mass as the total mass of the star decreases due to wind mass loss. This trend continues until the end of core helium burning, when core convection stops. When helium is depleted in the core, helium burning is ignited in a shell above the core. Afterwards, this particular evolutionary sequence goes through carbon burning in a series of convective flames that appear at subsequently higher mass coordinates, before igniting neon in the core. The layers enriched by helium burning ashes are never exposed by mass loss, and thus this model has $\mathrm{Y}_{\textrm{surf}} = 1-Z_{\textrm{init}}$ and a relatively high surface nitrogen abundance throughout its entire evolution.

In comparison, an evolutionary sequence with the same initial mass and metallicity of 0.04 (upper right panel in Fig. \ref{fig:kips}) loses mass at a higher rate, and thus its convective helium-burning core also decreases in mass at a faster pace. During core helium burning, the layers depleted of nitrogen and enriched in carbon and oxygen are exposed. At this point, the wind transitions from the WN regime to the WC regime, resulting in an increase in mass loss rate which also accelerates the retreat of the convective core. Subsequent episodes of nuclear burning in the core occur similarly, but the final mass is smaller, and the final surface composition is different, having a lower helium content, and with the ashes of helium burning also present.

To illustrate the effect of initial helium mass, we also show an evolutionary sequence with initial  mass of 20$\mso$ and metallicity of 0.01 (corresponding to a star with ZAMS mass of about $52\mso$, see lower left panel in Fig. \ref{fig:kips}). This model experiences the transition between mass loss regimes early in its evolution, similar to the 8$\mso$ evolutionary sequence at a metallicity of 0.04. However, it has a different core evolution due to its higher mass. Carbon burning in the core occurs radiatively, and as it progresses, the burning region moves to higher layers of the star, until it settles at the base of a convective zone that encompasses almost the entire carbon-oxygen core. This is common for stars of high mass, and affects the core structure, likely leading to the formation of a BH instead of a SN explosion \citep{2001NewA....6..457B,2014ApJ...783...10S}.

At a metallicity of 0.04, the evolutionary sequence with an initial mass of 20$\mso$ (see lower right panel in Fig. \ref{fig:kips}) has a stronger mass loss and experiences a wind-regime transition at an earlier evolutionary phase than its lower-mass counterpart. Due to its smaller mass at the end of core helium burning, it experiences carbon burning in a series of convective flames like the lower mass examples, linked to a higher probability of ending in a successful SN explosion at the end of its evolution (cf. \citetalias{aguilera-dena2021}).

As illustrated by these examples, populations of stripped-envelope stars in different metallicity environments will appear different. Stripped-envelope stars in high metallicity environments will spend a longer fraction of their lives with their carbon- and oxygen-rich layers exposed, they will have lower final masses compared to populations in lower metallicity environments, and the population of SNe they produce will be different, both in terms of total SN rates, in relative rate of different SN sub-types and in observable SN properties (cf. \citetalias{aguilera-dena2021}).

\subsection{Lifetimes and duration of evolutionary stages}\label{sec:lifetimes}
Hydrogen-free classical WR stars can be classified as either WN, WC or WO type. Each of these types corresponds to an evolutionary stage, and they are characterised by different surface abundance patterns and luminosity-mass ratios. Their occurrence, properties and relative rate is known to be different in environments of different metallicities, and has been used as a testbed of massive star evolution \citep[e.g.][]{1998ApJ...505..793M,2002ApJ...580L..35M,2005A&A...429..581M,2007ApJ...662L.107V,2012A&A...542A..29G,2017PASA...34...58E,2019Galax...7...74N,pauli}. The probability of observing a WR star either as a WN, WC or WO depends strongly on the duration of each of these stages. In this section, we analyse the duration of each stage in our helium star models according to their surface abundances and evolutionary stage, regardless of their surface optical depth. Whether their surface optical depth is large enough to be observed as WR type stars is discussed in Sect. \ref{sec:tau_analysis}.

\begin{figure}
\centering
\resizebox{\hsize}{!}{\includegraphics{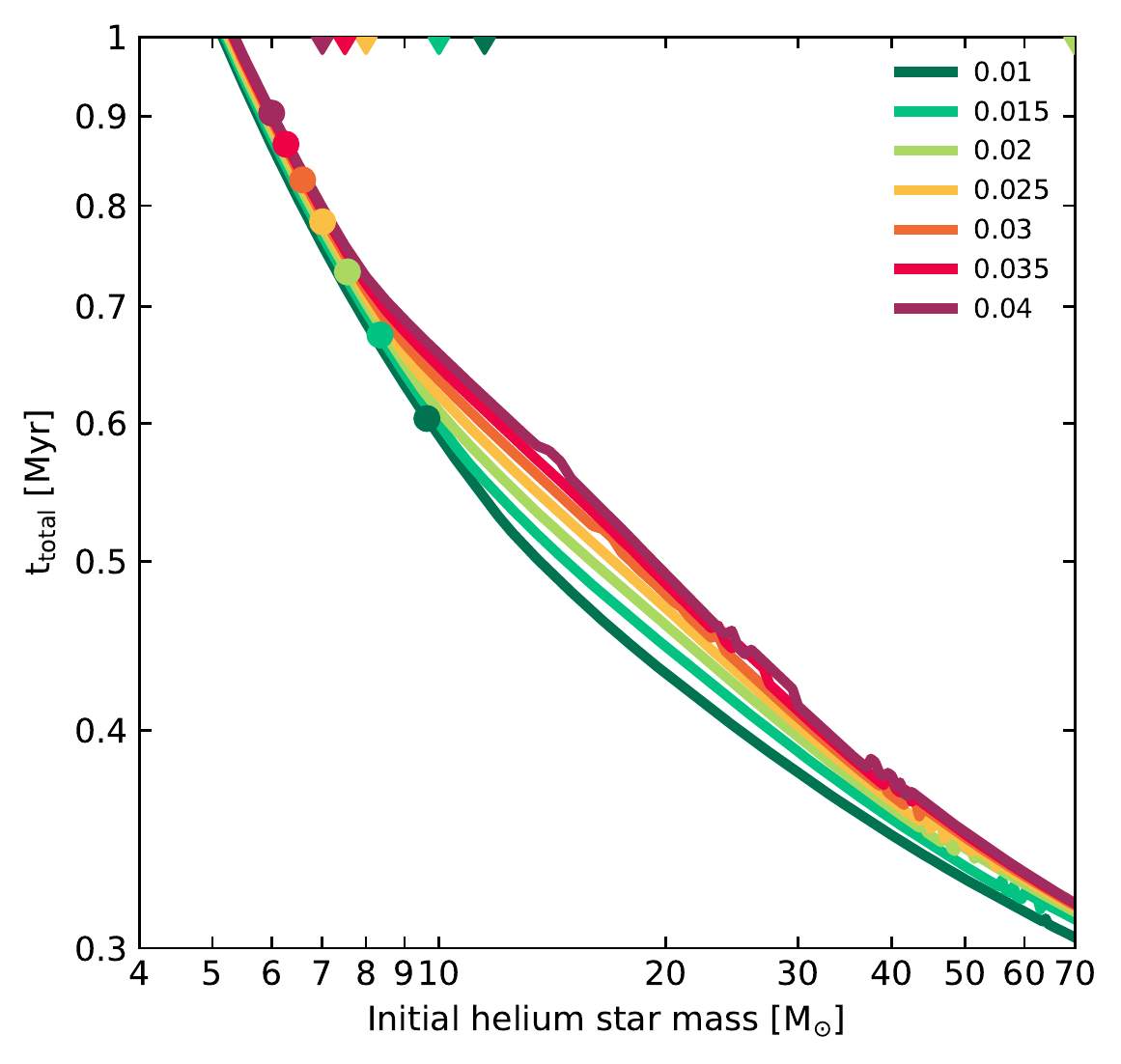}}
   \caption{Total lifetime of helium stars, from the start of helium core burning to core-collapse, as a function of initial helium star mass. Different coloured lines represent different metallicities. The downward pointing triangles at the top indicate the lowest initial helium star mass needed for a model to transition into the WC stage before the end of its evolution. The dots on each line indicate the minimum mass at which stripped-envelope stars may be classified as WN stars, obtained Eq. \ref{eq:fit_wn}, and using the mass-luminosity relation of \cite{1989A&A...210...93L}. Helium stars below this limit do not have optically thick WR winds, and have overestimated mass loss rates.}\label{fig:lifetimes}
\end{figure}

\begin{figure}
\centering
\includegraphics[width=8.1cm]{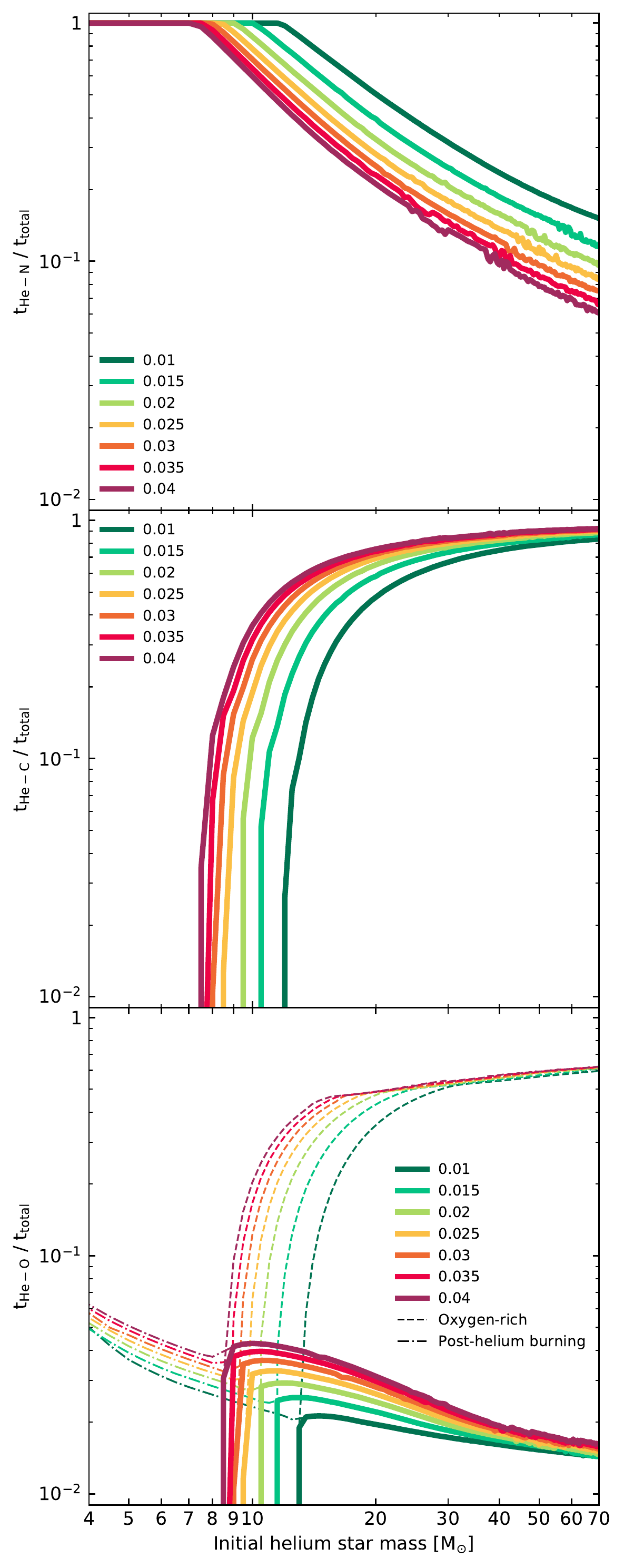}
\caption{Ratio between lifetimes of different evolutionary phases and total lifetime as a function of initial mass, for helium stars of different metallicities. The top panel shows the ratio between time spent with a nitrogen-rich envelope and total lifetime. The middle panel shows the ratio between time spent with a carbon-rich envelope and total lifetime. The bottom panel shows the ratio between lifetime spent with an envelope with surface oxygen abundance larger than 0.05 after core helium burning and the total lifetime. The dashed lines in the bottom panel indicate the ratios between the lifetime spent with an envelope with surface oxygen abundance larger than 0.05 and the total lifetime, and the dotted-dashed lines indicate the ratios between the lifetime after core helium depletion and the total lifetime. }\label{fig:types}
\end{figure}

The total lifetimes of our helium star models are shown in Fig. \ref{fig:lifetimes}, and available online\footnote{\url{https://zenodo.org/deposit/5747933}}. Similar to main sequence lifetimes, helium burning lifetimes of stripped-envelope stars are a monotonically decreasing function of initial mass. The lifetimes of models that do not expose the products of helium burning in their surfaces are well described by their helium burning nuclear timescale at the moment of stripping. The lifetime between core helium depletion and core collapse is only 2--6 \% of the total lifetime (see Fig. \ref{fig:types}, bottom panel). Models at lower metallicities are shown to have shorter lifetimes due to the fact that they retain a larger amount of mass at any given evolutionary stage, but the difference is limited to a few percent in this regime. Metallicity has a larger effect on helium star models that spend time with a carbon- and oxygen-rich surface (to the right of the circles in Fig. \ref{fig:lifetimes}) since they lose mass at a faster rate, and therefore decrease their convective helium-burning core mass early in their evolution. Helium star models that at some point have carbon- and oxygen-rich surfaces at the highest metallicity in our grid have lifetimes that are up to $\sim$ 14\% shorter than their low metallicity counterparts, particularly in the regime of initial helium star masses of ${10 - 20\,\mso}$.

In our simplified picture, where stellar models are assumed to be fully stripped of their hydrogen envelope at helium ignition, the duration of the different evolutionary phases depends on their total lifetime and the strength of their winds. The initial surface helium mass fraction of our models is given by $\mathrm{Y}_{\text{surf}} = 1 - Z_{\text{init}}$. The surface abundances of carbon, nitrogen and oxygen are set by CNO equilibrium, and are consequently initially nitrogen-rich, and carbon-and oxygen-poor. Therefore, all stellar models spend at least a fraction of their lifetime with nitrogen- and helium-rich envelopes (defined as $\mathrm{t}_\mathrm{He-N}$), and are thus candidate WN stars. Models that have helium burning ashes in their surface are candidate WC or WO type stars. We distinguish them by defining WC star candidates as those that are carbon-rich in the surface, and candidate WO stars as those that have a surface oxygen abundance of at least 0.05 and have finished core helium burning, after which their luminosity-mass ratio sharply increases. The duration of the carbon-rich phase (defined as $\mathrm{t}_\mathrm{He-C}$) is set by how quickly stars become stripped of their helium-nitrogen envelope. The duration of the oxygen-rich phase after core helium depletion (defined as $\mathrm{t}_\mathrm{He-O}$) is set by how quickly stellar models reach the threshold surface oxygen abundance in the lower mass regime, and by the remaining lifetime after core helium depletion in the higher mass regime. The fraction of the total lifetime that our models spend in each of the phases is summarised in Fig. \ref{fig:types}. 

The minimum mass at which models spend a fraction of their total lifetime as both WC and WO type star candidates decreases sharply with metallicity. The fraction of the lifetime that stellar models of the same initial mass spend in these more advanced stages is also an increasing function of metallicity. Combined with the increasing total lifetime, the relative number of WC and WO stars is expected to be strongly dependent on metallicity.

\subsection{Minimum luminosities of carbon- and oxygen-rich stripped-envelope stars}\label{sec:lums}

The minimum luminosity at which stripped-envelope stars are produced depends exclusively on the efficiency of their formation channel, and its dependence on ZAMS mass, binary fraction and metallicity. Whether such stars are WN or transparent wind helium stars depends on their surface optical depth.

Similarly, the emergence of carbon- and oxygen-rich stripped-envelope stars will depend on the efficiency of mass loss in stripped-envelope stars to lose their nitrogen-rich layer. Therefore, the minimum luminosity of carbon-rich and oxygen-rich stripped-envelope stars will depend on the metallicity of their environment.

\begin{figure*}[ht!]
\centering
\resizebox{\hsize}{!}{\includegraphics{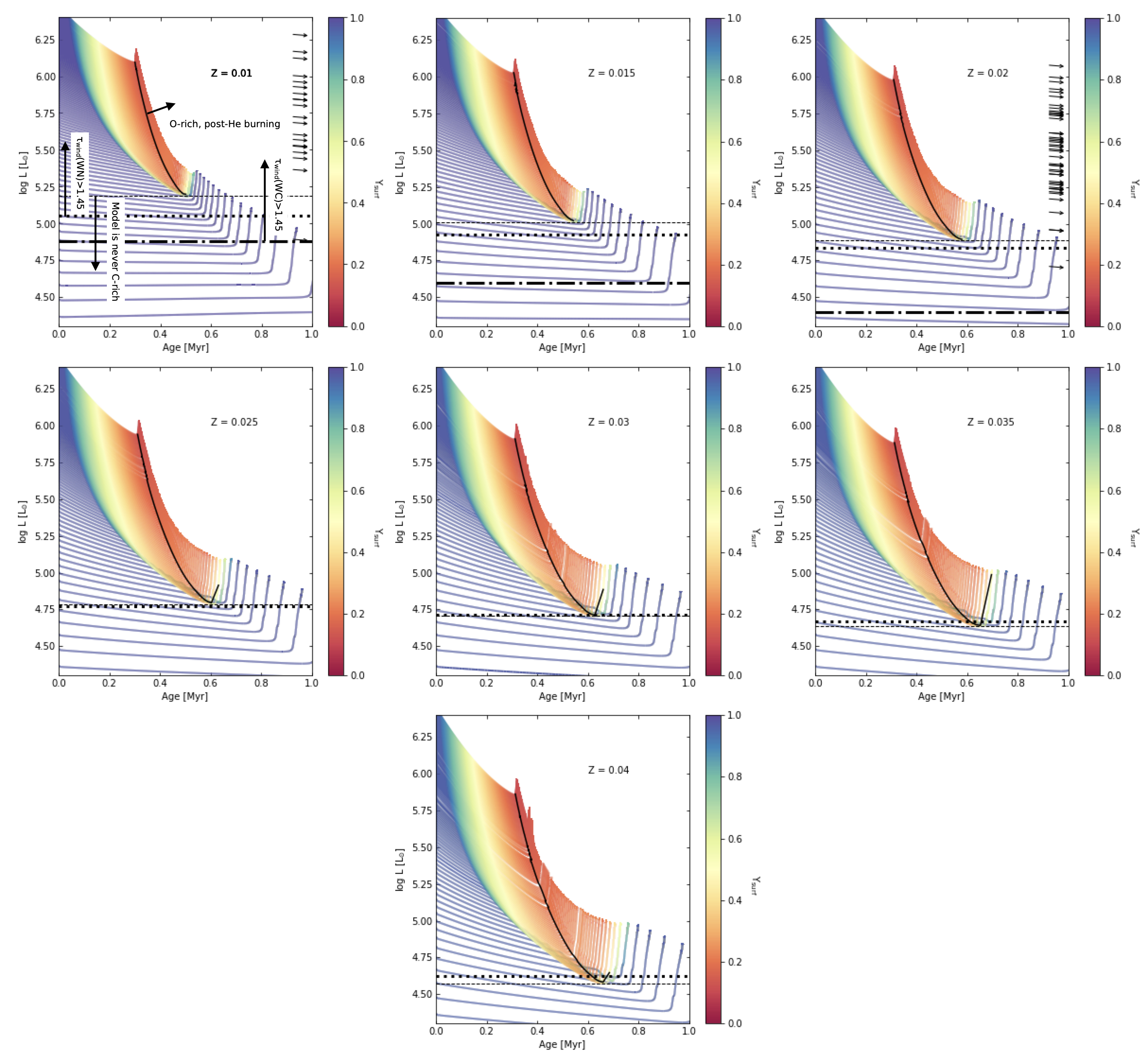}}
\caption{Evolution of surface luminosity as a function of age, between the He-ZAMS and core carbon depletion, for helium star models with different initial masses, coloured by surface helium abundance. Each panel represents a set of models with different metallicity, going from 0.01 (top left) to 0.04 (bottom). Arrows on the right hand side of the panel with Z=0.01 represent the luminosities of observed WC stars in the LMC, inferred from the sample of \cite{2001MNRAS.324...18B}, assuming a bolometric correction of $\mathrm{M}_{\mathrm{bol}} = 4.5$ \citep[e.g.][]{1989ApJ...347..373M,1994A&A...287..835S}, and have typical uncertainties of 0.3 dex. Arrows on the right hand side of the panel with Z=0.02 represent the luminosities of observed WC stars in the Galaxy, taken from \cite{2019A&A...621A..92S}. The thin, horizontal dashed lines represent the minimum luminosity at which carbon-rich models are formed at each metallicity, $\mathrm{L}^{\mathrm{evo}}_{\mathrm{min,WC}}$. The solid, black lines indicate the moment at which stellar models end core helium burning, and have surface oxygen abundance larger than 0.05. Horizontal, thick, dotted lines indicate the minimum luminosity at which the optical depth of nitrogen-rich, stripped-envelope stars is expected to be larger than 1.45, according to Eq. \ref{eq:fit_wn}. Horizontal, thick, dashed-dotted lines indicate the minimum luminosity at which the optical depth of carbon-rich, stripped-envelope stars is expected to be larger than 1.45, according to Eq. \ref{eq:fit_wc}. \label{fig:luminosities}}
\end{figure*}

Figure \ref{fig:types} shows that stars with an enhanced carbon and oxygen surface abundance will typically have larger initial masses than their nitrogen rich counterparts, and will therefore be found at preferentially higher luminosities. Carbon- and oxygen-rich stellar models are only produced above a threshold luminosity, $\mathrm{L}^{\mathrm{evo}}_{\mathrm{min,WC}}$. Stellar models with oxygen surface abundance above 0.05 after core helium burning are produced above a threshold luminosity $\mathrm{L}^{\mathrm{evo}}_{\mathrm{min,WO}}$. The value of these luminosities is determined from our models and is indicated in Fig. \ref{fig:luminosities}. Both limits decrease with increasing metallicity due to the stronger winds being more efficient at exposing the formerly convective cores of our helium star models.

As shown in Fig. \ref{fig:kips}, massive models lose mass at a rate large enough that their convective helium-burning core will decrease in size. This leads to a gradual luminosity drop during helium burning, as shown in Fig. \ref{fig:luminosities}. The drop in luminosity is larger at higher metallicities, and faster for models with carbon-rich surfaces that transition to WC mass loss rates. Models with lower masses have a more steady luminosity during their evolution, as their winds are not strong enough to decrease their mass significantly, and they do not transition to WC type mass loss.

Much like in models of stars with hydrogen rich envelopes, helium star models will experience an increase in luminosity a few thousand years before the end of their evolution, caused by the ignition of helium shell burning when helium is depleted in their core. The mass of the least massive model that has oxygen surface abundance above 0.05 determines the value of $\mathrm{L}^{\mathrm{evo}}_{\mathrm{min,WO}}$.

For models with luminosities below the minimum luminosity at which helium burning products are exposed at the surface, the final surface helium abundance corresponds to the initial one. Above this limit, helium stars expose their formerly convective layers, and their helium surface abundance decreases steadily, following the smooth composition gradient that is left above the convective helium-burning core. In helium star models with intense enough mass loss, a saturation value of around $\mathrm{Y}_\mathrm{surf} \sim $0.3 is reached after helium burning.

\begin{figure}
\centering
\resizebox{\hsize}{!}{\includegraphics{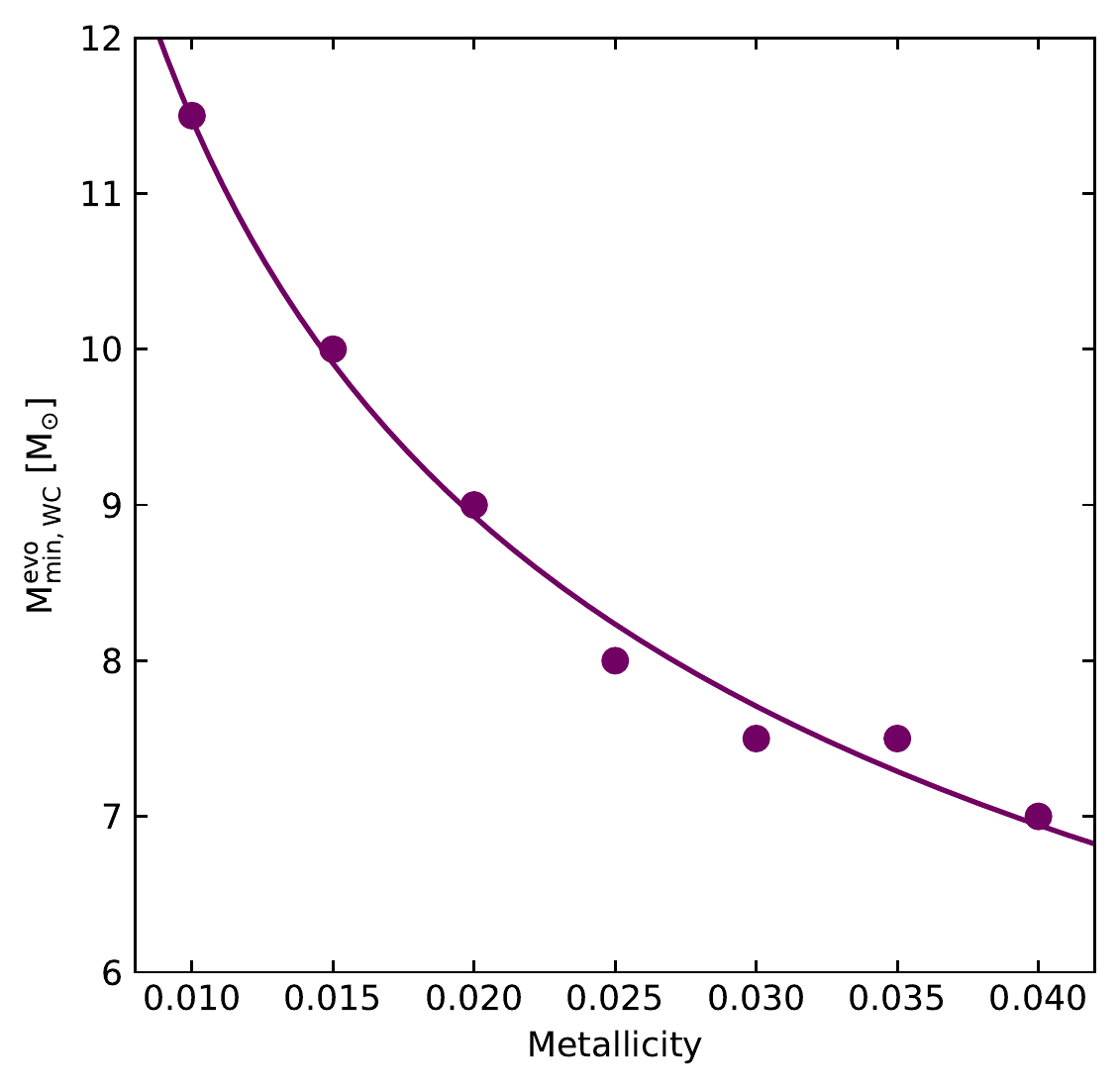}}
   \caption{Minimum initial helium star mass $\textrm{M}_{\textrm{min,WC}}^\textrm{evo}$ at which a stellar model exposes its carbon and oxygen enriched layers (potential WC type stars), as function of metallicity (dots). The minimum initial helium star mass can be described as a power-law, shown as a solid line (see Eq. \ref{eq:m_min_fit}).}\label{fig:m_min}
\end{figure}

Figure \ref{fig:m_min} shows the minimum \textit{initial} helium star mass, above which our models are found to expose the products of helium burning in their surface before core collapse. This minimum mass decreases with increasing metallicity, due to the increase in mass loss rates in the WN stage. Since the mass loss rates also increase with increasing luminosity, which is a proxy of mass \citep[e.g.][]{1989A&A...210...93L,2011A&A...535A..56G}, helium stars that are massive enough to expose layers of their formerly convective helium-burning core will spend a fraction of their lifetime as WC candidates. This fraction will increase the more massive they are. Therefore, as illustrated in Fig. \ref{fig:types}, for a fixed initial mass, the fractions of stellar lifetime spent as a WC and a WO candidate increase as a function of metallicity. This implies that WC and WO stars ought to be more common in environments of higher metallicity. The value of the transition mass can be fitted by a power-law. This fit is given by
\begin{equation}\label{eq:m_min_fit}
    \textrm{M}_{\textrm{min,WC}}^\textrm{evo} = 8.93 
    \left(\frac{\textrm{Z}}{0.02}\right)^{-0.362} \ \mso
\end{equation}
and is shown in Fig. \ref{fig:m_min}.

\section{Wolf-Rayet stars versus transparent wind stripped-envelope stars}\label{sec:tau_analysis}

In this section, we contrast the minimum luminosity of WR stars of different types --calculated employing the criterion based on the optical depth of the stellar wind, described in Sect.~\ref{sec:tau_method}-- to the outcome of the models presented in Sect. \ref{sec:results}. In Sect. \ref{sec:min_wn} we discuss what type of WR star can be expected to be the least luminous as a function of metallicity. In Sect. \ref{sec:min_wc} we discuss the minimum luminosities of WN, WC and WO stars, respectively, in the context of the results of our models. Finally, in Sect. \ref{sec:observations}, we discuss how our results compare to observed samples of WR stars in the local Universe.

\subsection{The least luminous Wolf-Rayet stars as a function of metallicity}\label{sec:min_wn}

The luminosity below which WR stars transition into transparent wind stripped-envelope stars can be estimated as a function of metallicity via the optical depth at their surface, as outlined in Sect. \ref{sec:tau_method}. Helium- and nitrogen-rich stripped-envelope stars with luminosities below the minimum WN luminosity may evolve with lower mass loss rates than those employed in our models (see Sect. \ref{sec:intro}). If they are stripped by a binary companion and have no further interactions after being stripped, their final masses are expected to be closer to their mass at the moment of stripping, their luminosities are expected to decrease less during their evolution, and their lifetimes are expected to be shorter.

Stellar models with helium- and nitrogen-rich surfaces in our grids with luminosities larger than $\mathrm{L}_\mathrm{min,WN}^\mathrm{tau}$ are therefore representative of WN stars. As shown in Fig. \ref{fig:lums_both}, in the metallicity range covered by our models, we find that the minimum luminosity above which stripped-envelope stars with nitrogen-rich helium envelopes are likely classified as WN stars is larger than the minimum luminosity above which stripped-envelope stars with carbon- and oxygen-rich envelopes are likely classified as WC stars (i.e., $\mathrm{L}_\mathrm{min,WN}^\mathrm{tau} > \mathrm{L}_\mathrm{min,WC}^\mathrm{tau}$). This implies that the lowest luminosity WR that \emph{can} exist in these environments is of WC type, regardless of whether they are formed by wind mass loss or not.

By extrapolating Eqs. \ref{eq:fit_wn} and \ref{eq:fit_wc} to higher metallicities, we find that this trend persists since the minimum luminosity of WC stars has a steeper metallicity dependence than the minimum luminosity of WN stars. However, the roles become inverted below metallicities of 0.0014, where according to our best fit models $\mathrm{L}_\mathrm{min,WN}^\mathrm{tau} = \mathrm{L}_\mathrm{min,WC}^\mathrm{tau}$. We predict that in environments with metallicities below this value, the least luminous WR star that can exist will be of WN type instead. 

The considerations made to derive the minimum luminosities of WN and WC stars carry no information about how they are formed, but rather just give the luminosity ranges where we could potentially observe stripped-envelope stars as WR stars. Whether they \emph{are formed or not} depends first of all on the probability of stripping hydrogen-rich stars, and subsequently on the intensity of their winds.

If the metallicity dependence of WN, WC and WO winds can be extrapolated to environments of low metallicity, our findings imply that both WN and WC stars become very rare in such environments, as the minimum luminosity increases quite quickly with decreasing metallicity, and they will only originate in very massive stars with high luminosities. Furthermore, the least luminous WR star that can exist in environments with metallicities below about 0.0014 (according to Eqs. \ref{eq:fit_wn} and \ref{eq:fit_wc}) will be a WN star, as opposed to more metal rich environments where the least luminous WR stars that can possibly exist are of WC type.

As an example, in a galaxy like I Zwicky 18, with a metallicity of 0.0002 \citep{2015A&A...581A..15S}, we predict that the minimum luminosity of a at which a nitrogen-rich stripped-envelope star has $\tau \leq 1.45$ is about $1.8 \times 10^{6} \lso$, and for a carbon-rich stripped-envelope star it is about $9.5 \times 10^{6} \lso$, corresponding to minimum masses of about 47 $\mso$ and 147 $\mso$, respectively. 

As a final note, we emphasise that the metallicity dependence of the minimum luminosities of WN and WC stars found through the optical depth model presented in Sect. \ref{sec:tau_method} have the empirical mass loss rates of WR stars as an input, so we caution that the behaviour of these quantities might change as our understanding of the mass loss rates of WR stars improves.

\subsection{Luminosity range of WC and WO type stars as a function of metallicity}\label{sec:min_wc}

\begin{figure*}
\centering
\resizebox{\hsize}{!}{\includegraphics{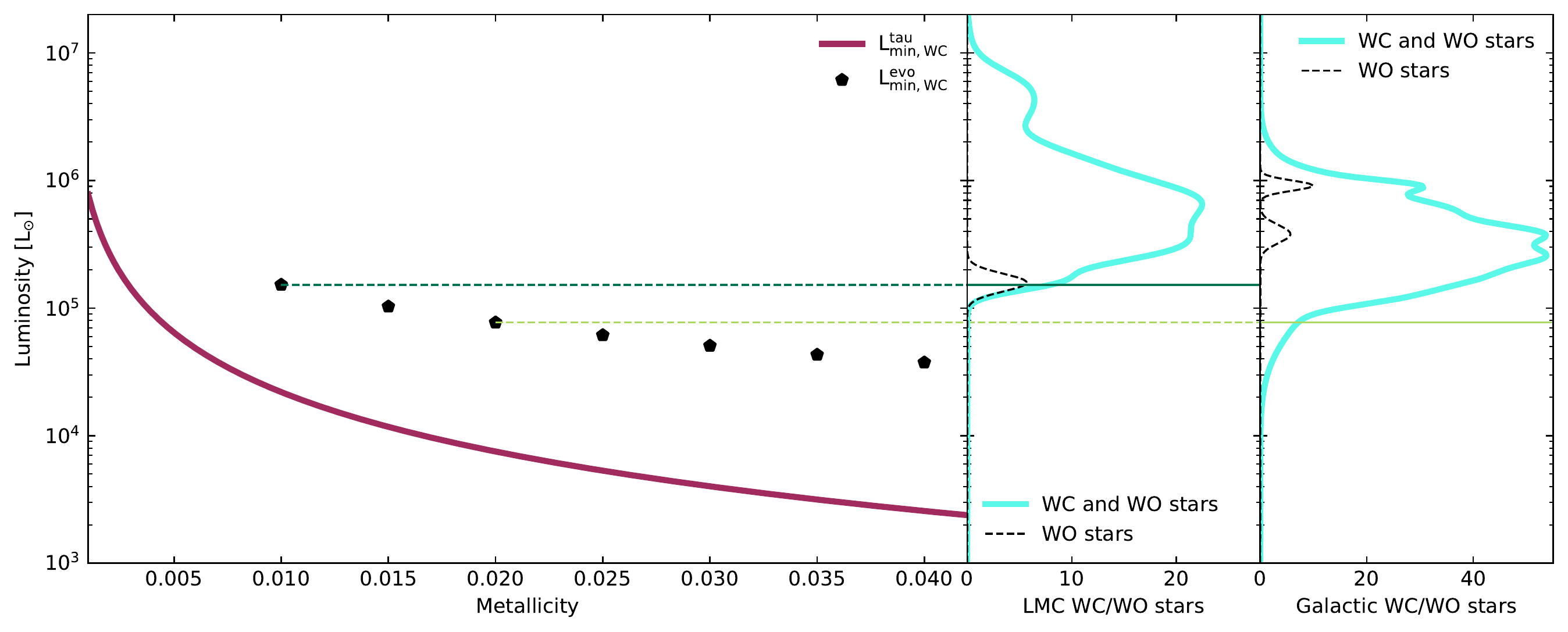}}
\caption{The solid line in the left panel shows the minimum luminosity at which WC stars are predicted to be observable as a function of metallicity, according to Eq. \ref{eq:fit_wc}. The black pentagons indicate the minimum luminosity at which our stellar models expose their carbon- and oxygen-enriched layers. Panels to the right show the observed luminosity distribution of WC and WO stars in the LMC \citep{2014A&A...565A..27H} and in the Galaxy \citep{2019A&A...621A..92S}. Horizontal lines compare the minimum luminosity carbon- and oxygen-rich stars produced in our models at $Z=0.01$ and $Z=0.02$ with the observed distributions in the LMC and the Galaxy, respectively.
}\label{fig:LMmin}
\end{figure*}

The least luminous carbon-rich stripped-envelope models produced in our grids of evolutionary calculations are dictated only by the effect of mass loss. As shown in Fig. \ref{fig:LMmin}, these models have luminosities well above the luminosity at which carbon-rich stripped-envelope stars would transition into having optically thin winds. The minimum luminosity carbon-rich models in our grids have optical depths $\tau \gtrsim 3$, and more luminous models have even larger optical depths. Therefore, they are representative of WC stars. Since the mass loss rate of WC and WO stars we employ has the same functional form, we find that all of our oxygen-rich, core helium depleted models are representative of WO stars after core helium depletion as well.

Furthermore, we find that the luminosity ranges at which WC and WO stars are formed are similar. However, we expect that the latter are more common at low luminosities, since the remaining lifetime between helium core depletion and core collapse decreases as helium star mass increases (see Figs. \ref{fig:types} and \ref{fig:luminosities}). Models representative of WC stars, on the other hand, spend a longer fraction of their lifetime as WC stars with increasing mass, so the luminosity distribution of WC stars is expected to result from a competition between the formation probability of a stripped-envelope stars of a given mass (which is partly determined by the IMF, and decreases as mass increases), and the amount of time they spend in the WC phase (which increases with increasing helium star mass).

WC and WO stars with luminosities lower than those found in our models can potentially be formed if there are additional channels that can strip helium stars of their nitrogen-rich helium envelopes, making their surfaces carbon-rich and nitrogen-depleted, such as subsequent episodes of binary mass transfer. Alternatively, they can be a consequence of mixing processes that either make the convective helium burning cores of stripped-envelope stars larger than what we find in our models, or efficiently exchange material between the envelope and the convective helium burning core.

The least luminous carbon-rich models produced in our grids reach luminosities below the minimum luminosity of WN at metallicities of about 0.03 and above. This implies that WC stars that are less luminous than the least luminous WN star \emph{are predicted to be formed via wind mass loss only in these environments}.

In the lower metallicity regime, since we expect that the minimum luminosity at which carbon- and oxygen-rich stripped-envelope stars can be observed as WC stars will increase, WN stars may be capable of forming carbon-rich and nitrogen-poor helium stars with transparent winds via wind mass loss, but further modelling of stripped-envelope stars in these environments is required to confirm this conjecture.

We note that evolutionary tracks in our grids that expose the products of helium burning during their evolution always have luminosities above the minimum luminosity of WN stars inferred with the wind optical depth method at all metallicities. This means that our numerical modelling employing the \cite{2017MNRAS.470.3970Y} mass loss rates is correctly applicable to all evolutionary tracks that spend time in the WC regime, as they do not transition into the regime where we expect their mass loss rates to be lower. We emphasise that the minimum luminosity of WN stars is independent of the evolutionary state of our models, whereas the minimum luminosity of WC in our grids of evolutionary calculations is set by the evolutionary history of our models, which decrease in luminosity during their evolution as a consequence of mass loss (see Fig. \ref{fig:luminosities}), and does not correspond to the initial luminosity of our models.

\subsection{Comparison with observations}\label{sec:observations}

Figure \ref{fig:LMmin} shows the luminosity distribution of WC stars in the LMC and in the Galaxy, compared to the predicted minimum luminosities from our optical depth model, and those achieved in our evolutionary calculations. As shown in this Fig. \ref{fig:LMmin} (and in Fig. \ref{fig:luminosities}), all of the WC and WO stars observed in the LMC and in the Galaxy are above the minimum luminosity we predict using our optical depth model, and the bulk of their populations are within the luminosity range in which they are formed by our models. This implies that most of them are consistent with having been formed from a WN star that lost its nitrogen-rich helium envelope due to winds, and do not require an additional source of mass loss to be formed. In the LMC, the fact that the only WO star known is close to the lower luminosity at which such stars are produced in our models is also consistent with the fact that the least luminous WO stars also have the relatively longest lifetimes.

In our Galaxy, however, a tail in the luminosity distribution of WC stars populates the region where we predict WC and WO stars \emph{could} exist according to our optical depth model, but where no candidates are found in our evolutionary calculations. The tail of the luminosity distribution of WC stars in the Galaxy contains few stars, and the luminosities determined for these stars have significant uncertainties. A possible explanation is that stars that populate this tail may be found in environments within our Galaxy with higher metallicity, but if the tail is significant for the commonly adopted Galactic metallicity value, it may imply that WC and WN type winds alone are unable to produce WC stars in the luminosity regime of the least luminous observed WC and WO stars. This might either imply that stars below this limit are formed by a different mechanism, such as case BB mass transfer, that their luminosities are underestimated, or that either WN or WC mass loss rates are higher than what we employed in our models. The least luminous star in the Galactic sample is WR 119, with $\log \ \mathrm{L} / \lso = 4.7^{+0.25}_{-0.2}$. \cite{2019A&A...621A..92S} noted that it is the least luminous object in their sample, but the uncertainty in its parallax is large, implying that the real uncertainty in its luminosity could be up to 0.3 dex, still well below $\mathrm{L}^{\mathrm{evo}}_{\mathrm{min,WC}}$. However, this object is located toward the center of the Galaxy, where metallicity is potentially higher than in the Solar Neighbourhood.

Having a larger sample of observed WR stars, in a larger metallicity range, and a more stringent determination of their distances and parameters will help settle the question of whether or not transparent-wind carbon-rich stars can be formed via wind mass loss (as opposed to case BB mass transfer, see \citealt{2013ApJ...778L..23T}), as well as whether or not additional channels for forming WC and WO stars are necessary to populate the gap between the least luminous stars of these classes produced by WN mass loss, and the least luminous carbon-rich stripped-envelope stars.

Recent work has been made to predict the observational characteristics of helium stars with optically thin winds through spectral modelling \citep{2018A&A...615A..78G}, and this theoretical work has inspired observational campaigns that have produced candidate transparent wind stripped-envelope stars. These pursuits will provide a complementary perspective on the nature of stripped-envelope stars, by giving us a glance at the population of stripped-envelope stars of a lower mass. Information about them will constrain the stripping efficiency of stars as a function of mass, hopefully adding the missing pieces of the puzzle.

As a final note, we highlight that the existence of a minimum luminosity limit for WR stars has been established observationally \citep{2020A&A...634A..79S}, and therefore the results and discussion we put forward are independent of the model we use to calculate the minimum luminosity of WR stars.

\section{Wolf-Rayet populations in the Local Group}\label{sec:populations}

\begin{figure}
\centering
\resizebox{\hsize}{!}{\includegraphics{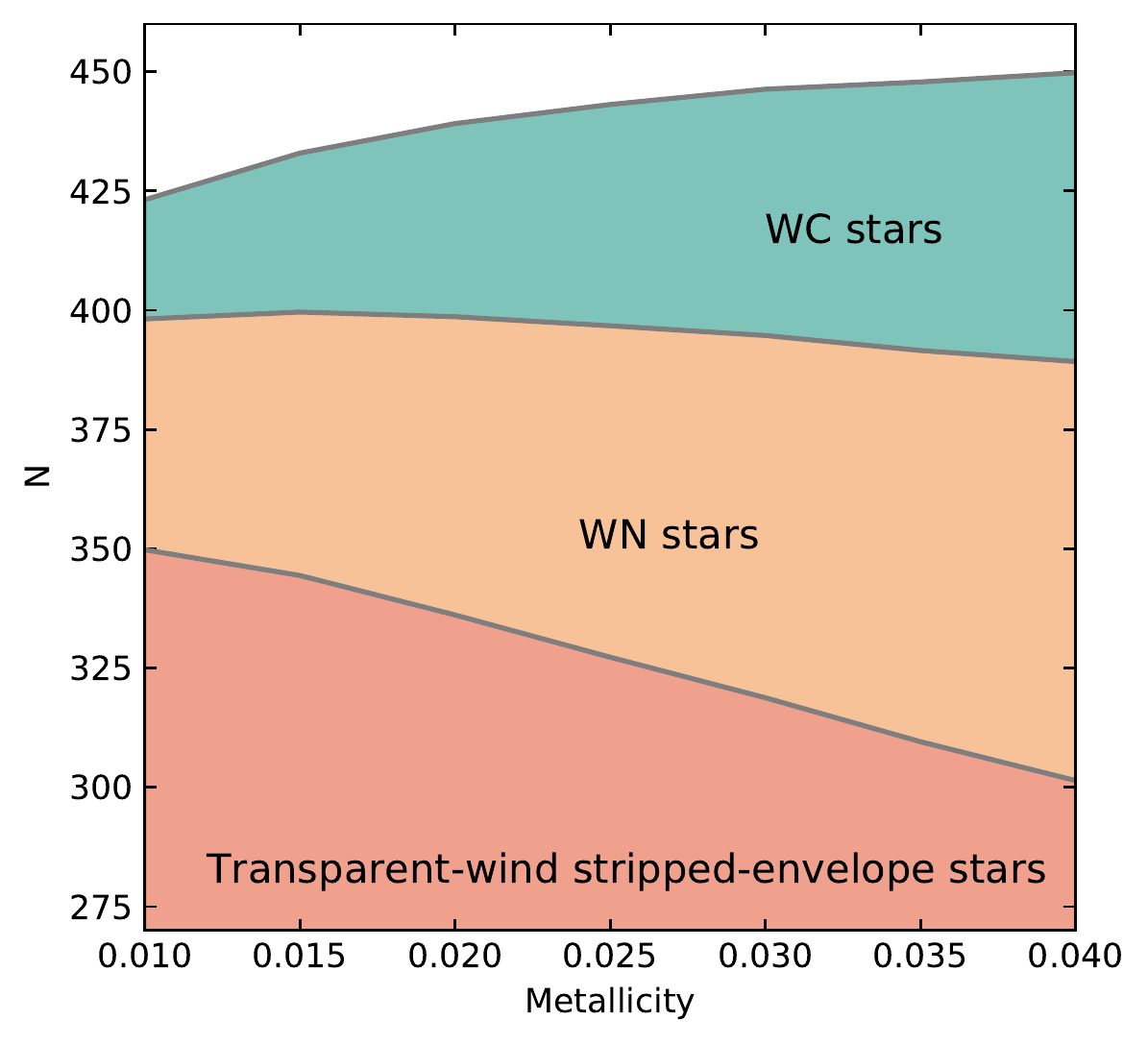}}
   \caption{Numbers of transparent-wind stripped-envelope stars,
   WN stars and WC stars obtained from our population models are shown as a function of metallicity. The numbers are normalised so that the population with metallicity of 0.01 has 25 WC stars, matching the population of the LMC \citep{2018ApJ...863..181N}.}\label{fig:nums}
\end{figure}

\begin{figure}
\centering
\resizebox{\hsize}{!}{\includegraphics{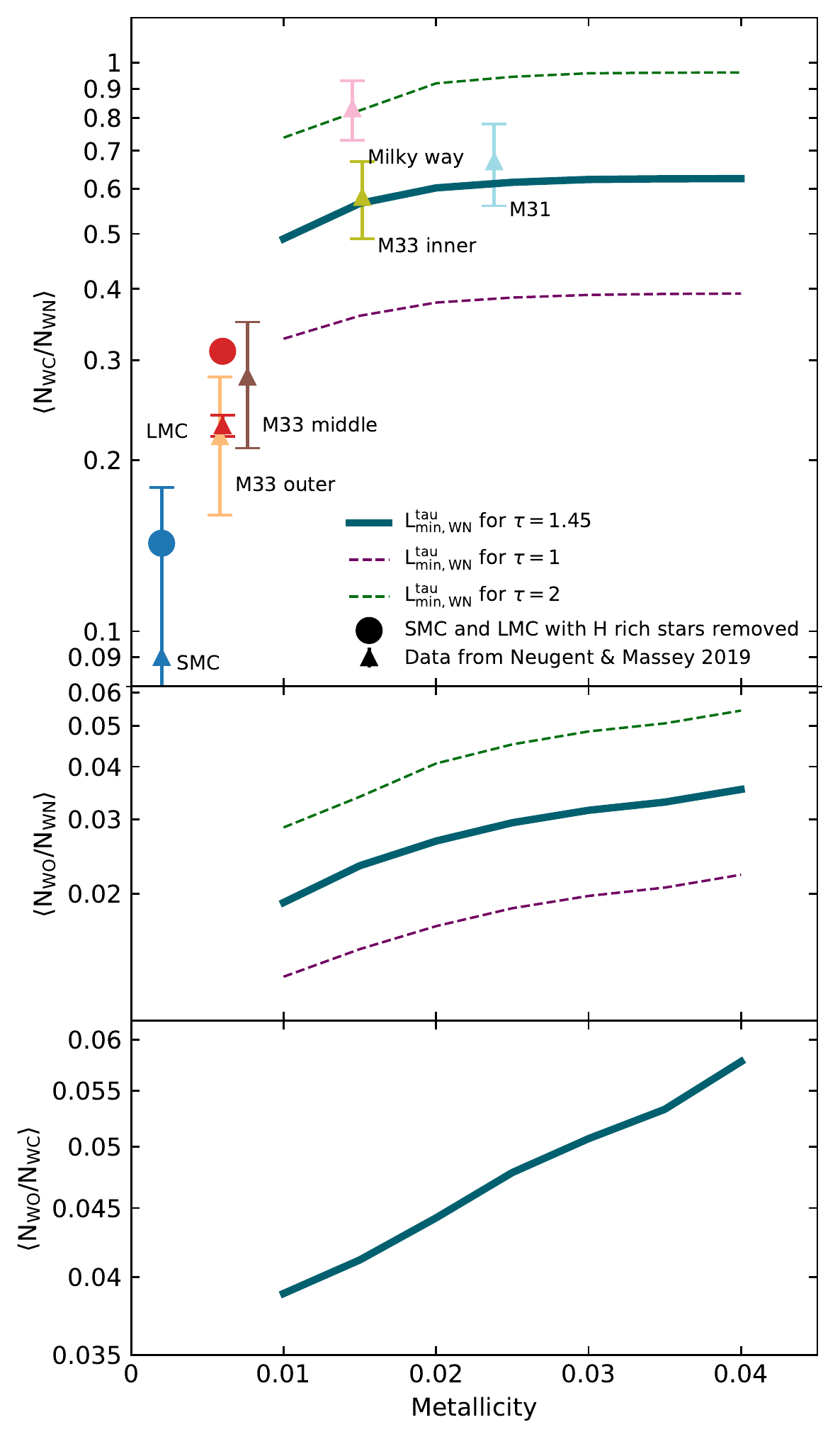}}
\caption{The average number ratio of WC to WN stars (top panel), WO to WN stars (middle panel) and WO to WC stars (bottom panel) as a function of metallicity, as predicted from our models, are represented by solid lines. This quantity strongly depends on the minimum mass for a star to be observable as a WR. The different lines in the top and middle panel correspond to employing different values of the minimum mass of WN stars, found adopting different values of wind optical depth for the transition between WN stars and transparent wind stripped-envelope stars. The solid line corresponds to the fit to observations in Eq. \ref{eq:fit_wn}, adopting $\tau = 1.45$ (blue). Dashed lines are calculated adopting $\tau = 1$ (purple) and $\tau = 2$ (green). Data shown in triangles correspond to the observed ratios of \citep{2019Galax...7...74N}, and correspond to lower limits due to the inclusion of hydrogen rich WN stars. Data shown in circles for the LMC and SMC correspond to the same ratio, with WN stars with $X_H>0.3$ removed (according to the analyses of \cite{2014A&A...565A..27H} for the LMC, and \cite{2015A&A...581A..21H,2016A&A...591A..22S} for the SMC).} \label{fig:wc_wn_ratio}
\end{figure}

Using the method described in Sect. \ref{sec:methods_popsynth} we create simplified synthetic populations of stripped-envelope stars. We compute the number of transparent wind stripped-envelope stars with initial masses above $3 \mso$, WN stars and WC stars as a function of metallicity, presented in Fig. \ref{fig:nums}. We normalise these so that the number of WC stars in the population with metallicity of 0.01 matches the 25 known WC stars in the LMC. We then compute the average number ratios of WC to WN stars, $\langle \text{N}_{\text{WC}}/\text{N}_{\text{WN}}\rangle$, WO to WN stars, $\langle \text{N}_{\text{WO}}/\text{N}_{\text{WN}}\rangle$, and WO to WC stars, $\langle \text{N}_{\text{WO}}/\text{N}_{\text{WC}}\rangle$, as a function of metallicity. These ratios are shown as thick, blue lines in Fig. \ref{fig:wc_wn_ratio}. In these calculations, it is assumed that the star formation rate, binary fraction, and stripping probability are the same at any given mass, and thus their uncertainties cancel out when calculating number ratios of different types of WR stars.

As shown in Fig. \ref{fig:nums}, the total number of stripped-envelope stars increases slightly as a function of metallicity. This is mainly due to the fact that, as metallicity increases, the lifetime of a stripped-envelope star of a given initial mass increases as well (see Fig. \ref{fig:lifetimes}). The populations are dominated by transparent-wind stripped-envelope stars at every metallicity, followed by WN type stars, then WC stars and finally WO type stars. Qualitatively, this is in agreement with the known populations of WR stars, where WN type stars represent the majority of WR stars, whereas WC type stars are progressively more rare, and WO type stars are very uncommon.

The decrease in the number of transparent-wind stripped-envelope stars, and the corresponding increase in the number of WN type stars as a function of metallicity, is caused by the decrease in the minimum luminosity of WN stars. As shown in Fig. \ref{fig:nums}, the number of transparent-wind stripped-envelope stars exceeds the number of WN stars by a factor of 8 in the lowest metallicity regime in our simulation, and by a factor of about 3 at the highest metallicity.

As shown in Figs. \ref{fig:nums} and \ref{fig:wc_wn_ratio}, the number of WC type stars relative to WN type stars increases as a function of metallicity. This is a consequence of two effects: The first of them is that, as metallicity increases, models with the same initial helium star mass spend more time as WC stars, as shown in Fig. \ref{fig:types}. The second is the decrease of the minimum initial helium star mass above which WC stars are produced in our evolutionary calculations. Both of these phenomena are consequence of the higher mass loss rates during the WN phase with increasing metallicity. However, the fact that this ratio increases as a function of metallicity is not trivial, since the minimum mass at which stellar models are classified as WN type stars also decreases with increasing metallicity, implying that lower mass helium stars, favoured by the IMF in our simple picture, are also produced in greater abundance. This is the reason why this ratio stalls above a metallicity of about 0.03 at a value of about 0.6.

Similarly, the increase in the number of WO type stars relative to both WN type and WC type stars follows from the fact that models with higher metallicity have higher mass loss rates, so that the oxygen-rich layers become exposed earlier in their evolution. Another fact that contributes to the increase of these two ratios in a population is that the time between core helium depletion and core collapse increases as a function of metallicity, as it is inversely proportional to the mass at the end of helium burning. Finally, this ratio also increases because the minimum mass at which a stellar model reaches the WO stage decreases as a function of metallicity, therefore becoming more favoured by the IMF.

As shown in Fig. \ref{fig:wc_wn_ratio}, we find that the number of WO type stars relative to both WN and WC type stars increases monotonically, without stalling at a fixed value. This is because the minimum mass at which stars enter the WO stage decreases faster than the minimum mass at which they become WC stars. Furthermore, as opposed to WC type stars, the least massive WO stars tend to also be the ones that spend the longest fraction of their lifetime in this phase. Therefore, on average, a larger proportion of WO stars in a population of WRs is a better indicator of metallicity than the number of WC stars relative to WN stars alone.

The number of WC and WO type stars relative to WN type stars are sensitive to the value of the minimum mass above which stripped-envelope stellar models would be classified as WN type stars. To show this, we repeat the calculations using a range of minimum WN masses that correspond to different assumptions on the optical depth at which the transition between WN and transparent wind stripped-envelope stars occurs. We adopt $\tau = 1$ and $\tau = 2$, shown with a dotted-dashed and a dashed line, respectively, in Fig. \ref{fig:tau}. The resulting ratios are shown in the purple and green dashed lines of Fig. \ref{fig:wc_wn_ratio}.

For all three choices of minimum WN mass, the numbers of both WC and WO type stars relative to WN type stars are found to increase as a function of metallicity with roughly the same slope. The increase is less than a factor 2 in all cases. This behaviour qualitatively reproduces the behaviour of the WC/WN ratio in observed populations \citep{2019Galax...7...74N}. The dependence on the choice of minimum mass is also steep due to the steepness of the IMF. Increasing the value of the minimum mass of WN stars excludes lower mass stars, favoured by the IMF, and more likely to spend time as WN stars than WC stars. Therefore, this choice will increase the number of WC stars at any given metallicity. 

On one hand, because of the steepness of the IMF, if the stripping probability is constant, most WN stars formed will be of relatively low mass, and cluster around the minimum WN luminosity. On the other hand, because the mass-luminosity relation of WN stars is also steep, a small variation in the minimum luminosity of WN stars results in a large variation in the number of WN stars. The combination of these two effects, as shown in Fig. \ref{fig:wc_wn_ratio}, has a very strong impact on the morphology of WR populations. We find that after a small variation in the adopted minimum mass of WN stars, this ratio varies by almost a factor of 2 regardless of metallicity. The variation that results from changing the lowest WN mass is larger than the variation that results from changes in metallicity by a factor 4, even though more stars evolve to the WC phase at higher metallicity, and that they spend a longer fraction of their remaining lifetime in this phase. 

This implies that the minimum luminosity at which stripped-envelope stars are observable as WR stars \emph{has a larger impact on WR populations} than the metallicity dependence of WR mass loss rates. Since all carbon- and oxygen-rich stars in the metallicity range of our calculations are expected to be observable as WR type stars, we expect that these ratios are also indicative of the minimum luminosity of WN type stars. Careful consideration of these findings is paramount to understand the evolution of WR stars and transparent-wind stripped-envelope stars in the Universe. Previous theoretical studies of WR stellar populations where this quantity has been computed have found that the stripping probability of main sequence stars, either single \citep[e.g.][]{2005A&A...429..581M} or in binaries \citep[e.g.][]{2017PASA...34...58E} is key to understanding this ratio, but we argue that including the behaviour of the minimum luminosity of WR stars as a function of metallicity can be just as important. 

The picture becomes more complicated in low metallicity environments, since other poorly constrained quantities, such as the minimum luminosity of WC stars, and the efficiency at which they are formed by WR winds become important, and complicate the matter further. However, we hope to see these considerations accounted for in future population synthesis models, to build a more accurate picture of the WR populations in different galaxies, as they have a very large impact in their environments.

The ratio of WO to WC stars does not depend on the wind optical depth model because all WC and WO candidates in our models are also found to have an optical depth that is significantly higher than 1.45. However, it is sensitive to the rest of the uncertainties that our models are subject to, such as the mass loss rates assumed, and to the definition of the boundary between WC and WO stars.

To assess the validity of our results, we compare our population with metallicity of 0.01 to the WR population of the LMC, which is known to be complete \citep{2018ApJ...863..181N}; and we compare the ratios of WC to WN type stars to the WC/WN ratios inferred by \cite{2019Galax...7...74N}. The observed values of $\log$ (O/H) + 12 from \cite{2019Galax...7...74N} were converted to metallicities assuming that the ratio of oxygen mass fraction to total metallicity is constant, and corresponds to the Solar value \citep{2009ARA&A..47..481A}, and that helium mass fraction increases linearly with metallicity from its primordial value to the value observed in the Sun \citep{2007ASPC..374...81P}. Observed values of $ \text{N}_{\text{WC}}/\text{N}_{\text{WN}}$ are made according to spectral class, which means that hydrogen rich WN stars are included in the WN sample. Therefore, these measurements represent a lower limit to the results in our calculations. For comparison, the values of WN/WC are shown for the SMC and LMC with hydrogen rich WR stars removed from the sample, according to the analyses of \cite{2014A&A...565A..27H}, and \cite{2015A&A...581A..21H,2016A&A...591A..22S}, respectively. 

We find that by imposing that the population with metallicity of 0.01 has 25 WC type stars, matching the number of WC stars in the LMC, the number of WN stars is underestimated. Our population model has 48.4 WN stars, while the LMC has a total of 124. However, out of the 97 WN analysed by \cite{2015A&A...581A..21H}, 74 of them have a surface hydrogen abundance of less than 0.3. The discrepancy might be due to the fact that the metallicity in the LMC is lower than 0.01, because our model excludes WN stars that originate from single stars, and because of the simplicity in our assumptions. Using this population model we predict that the number of transparent-wind stripped-envelope stars in the LMC should be around 350.

We find that the number ratios in Fig. \ref{fig:wc_wn_ratio} agree qualitatively with our findings, but not many environments have observed populations of WR stars in the metallicity regime of our calculations. At lower metallicities, the effect of remaining hydrogen in the envelope after binary interactions will likely become important in determining this quantity, so the comparison cannot be made directly with models similar to those presented here, performed at lower metallicity.

The number ratios involving WO stars are more difficult to assess in these populations due to an intrinsic rarity of WO stars. In the environments where observations have been performed, only very few WO stars are found. In the SMC, no WC type stars are known, and there is only one WO type star known. Using our models with metallicity of 0.01 as a proxy of the LMC, we predict that there should be between 2 and 3 WO type stars accompanying the 124 WN type stars known. This is in line with the two WO type stars known in this environment.

There are no known WO stars in M31 \citep{2012ApJ...759...11N} and M33 \citep{2011ApJ...733..123N}. M31 contains about 92 WN stars, which would imply that, according to our models, there should be between 2 and 3 WO type stars. Similarly, we would expect that the 45 WN stars in the inner part of M31 are accompanied by about 3 WO type stars. 

In the Milky Way the WR type ratios are difficult to determine since the observed WR population is not complete. Future observations of WR stars in our own Galaxy and others can contribute to measuring these ratios in environments of different metallicities, and this pursuit can potentially be used to put constraints on several aspects of stellar evolution.

\section{Discussion}\label{sec:discussion}

The results presented in the above sections were derived using 1D stellar evolution models of helium stars with metallicities in the range 0.01 -- 0.04.  Taking this approach, we are able to study the effects of WR mass loss on stripped-envelope stars, without the additional complications and uncertainties of the previous evolutionary stages. This range includes the metallicity of the Solar neighbourhood, but extends to environments of high metallicity, which have often been neglected in the literature. The metallicities included in this study are meant to explore the range where the amount of hydrogen remaining in the envelope after envelope stripping is more likely to be negligible. We aim to characterise the effect of metallicity dependent mass loss rates, as well as the metallicity dependence of the luminosity at which stripped-envelope stars are separated between WR stars and transparent-wind helium stars in the populations of WR stars; particularly through the ratio of WC to WN stars.

In the following subsections we address the limitations and uncertainties of the approach we have taken to obtain them, and compare them to previous models of stripped-envelope stars.

\subsection{Uncertainties in helium star evolution}
\subsubsection{Uncertainties in Wolf-Rayet mass loss rates}

The main effect of metallicity on the evolution of helium stars of equal initial mass, given the physical treatment that we employ, is due to the metallicity dependent winds. The wind mass loss rates used in this work were adapted from \cite{2017MNRAS.470.3970Y}. These were obtained from separate empirical studies of WN \citep{2014A&A...565A..27H} and WC stars \citep{2016ApJ...833..133T}, and calibrated to have matching clumping factors (namely ${D=4}$), which \cite{2017MNRAS.470.3970Y} found would help reconcile the lowest luminosity WC stars observed to our understanding of WR mass loss. 

The empirically obtained mass loss rates of WR stars used in our simulations were extrapolated beyond the luminosity and mass regimes at which the WR mass loss rate observations were carried out. Therefore, we caution that the results of our simulations, particularly those with masses near and below the obtained values of the minimum mass of WN stars at a given metallicity (indicated with diamonds in previous figures) have mass loss rates that may be overestimated.

Currently there is no consensus in the community as to what are the true mass loss rates of helium stars with different luminosities and compositions. Recently, 1D hydrodynamically consistent simulations of wind acceleration in WR stars have been carried out \citep{2020MNRAS.491.4406S,2020MNRAS.499..873S}. These studies suggest, contrary to what has been found empirically, that the mass loss rates of WR stars of similar parameters have a shallower dependence on metallicity than that determined from the observed sample of WN and WC stars. Furthermore, these studies suggest that the mass loss rates of WC type stars are in fact lower than those of WN type stars of similar luminosity and mass. If this is indeed the case, then the final masses, ejecta masses and BH masses predicted for stars in our sample that spend a fraction of their lifetimes as WC stars are also underestimated. Self-consistent theoretical calculations of WR mass loss rates are uncertain inasmuch as the temperatures (and therefore the properties of the wind-accelerating regions) are poorly constrained.

\subsubsection{Structure of massive stars after stripping}\label{sec:init_conds}

The initial condition of our simulations is an approximate representation of the state of stars that have lost their hydrogen-rich envelopes through winds, binary stable mass transfer or CE ejection. This approach has been employed in many studies of the late evolution of stripped-envelope stars \citep{1974ApJ...194..373A,1988PhR...163...13N,1995ApJ...448..315W,2015MNRAS.451.2123T,2016MNRAS.459.1505M,2017MNRAS.470.3970Y,2018MNRAS.481.1908K,2019ApJ...878...49W,2020ApJ...890...51E,2020ApJ...896...56W,2021MNRAS.505.4874H}, and has been found to be a powerful tool to study a complex stage in massive star evolution, that is very relevant in the study of Type I SNe, WR stars and gravitational wave sources.

Detailed simulations of binary systems that become stripped through stable mass transfer show that this interaction does not remove the entire envelope of the donor star in the binary system, and that a remaining hydrogen layer, even of a very low mass, can have significant consequences in the following evolution of these stars \citep{2019MNRAS.486.4451G,2020A&A...637A...6L,pauli}. Furthermore, stars that lose their hydrogen envelopes before the beginning of core helium burning cannot reproduce phenomena such as WN/WC stars, that are observed to be rich in both nitrogen and carbon, and require a mixing mechanism that likely arises from helium gradients that only occurs in convective helium-burning cores that grow in mass \citep{1991A&A...248..531L}.

Recent simulations of binary evolution at LMC metallicity \citep{pauli} show that interacting binary systems will retain a hydrogen-rich layer after stripping. At this metallicity, and using the mass loss rates from \cite{2017MNRAS.470.3970Y}, these layers are only expected to be present for $\sim$10 -- 20\% of the helium burning lifetime of the donor stars, depending on their mass, in short period systems. For the systems with the longest periods and for stars evolved in isolation, their models enter the hydrogen free WN stage after ${~50\%}$ of the helium core was burned. At higher metallicities the winds remove the hydrogen-rich layers more efficiently, therefore reaching the hydrogen-free WN stage earlier in their evolution.

Binary evolutionary calculations with different initial metallicities \citep{2021arXiv211110271K} suggest that this layer is efficiently removed in systems with Solar metallicity and above, suggesting that binary stripping is more efficient as metallicity increases. Therefore, we are confident that our models reproduce the properties of stars that have been stripped by Case A or Case B mass transfer, as well as successful CE ejections that occur soon after the ignition of hydrogen shell burning at the metallicities employed in our grids. Detailed binary stellar models are required to study stripped-envelope stars at metallicities lower than those in our grids, to properly account for the effect of the remaining hydrogen layer in the internal structure of stripped-envelope stars, as well as on their orbital evolution. Furthermore, as noted by \cite{2021A&A...645A...5S}, stars that interact earlier or later have different structures at the beginning of helium burning, and conclusions about such systems can therefore not be derived from our models.

However, the simulations of \cite{2019MNRAS.486.4451G} show that, if stars that retain a small amount of hydrogen in their envelopes do not lose winds as WR stars (but instead follow the \cite{2017A&A...607L...8V} prescription, made for stripped-envelope stars below the WR luminosity limit), then the remaining hydrogen is less likely to be removed, leading to the preferential formation of Type IIb SN progenitors, instead of Type Ib progenitors. If mass loss rates of stripped-envelope stars are indeed not large enough to efficiently remove the remainder of the hydrogen envelope after a binary interaction, then approximating them as pure helium stars from the beginning of their evolution may lead to overestimating their final mass, as well as overestimating the number of WC stars in a given population.

The structure of stars that have lost their hydrogen envelope through CE interactions is even more uncertain. CE evolution is typically modelled using the so-called energy formalism \citep{1984ApJ...277..355W}, which provides a method to calculate whether a binary system will merge, or survive a CE interaction and successfully eject the CE. In the latter case, it allows for the calculation of the final separation of the system. However, this formalism depends on an arbitrary choice of boundary between the stellar core, which will be the surviving part of the star after the interaction, and the envelope that will be ejected. Therefore, the surface composition of stars after the interaction is unknown. The range of possible choices and the consequences of variations in this choice have been discussed by several authors \citep[e.g.][]{2001A&A...369..170T,2011ApJ...730...76I,2016A&A...596A..58K,2021arXiv210714526V}, but no consensus has been found on the outcome of CE events. Recent studies of CE interactions \citep[e.g.][]{2019ApJ...883L..45F,2020arXiv201106630L} find that the final structure and separation of the binary components after the interaction is uncertain even in detailed simulations, and that the binary components might be subject to subsequent interactions during the stripped star's core helium burning phase. However, it is likely that most or all of the hydrogen envelope will be ejected if the event does not lead to a merger.

Single stars can also become stripped of their hydrogen envelope via their stellar winds. Helium stars below a certain mass are probably only formed due to binary interactions as their winds during the main sequence are unlikely to be strong enough to strip them of their hydrogen envelope. However, this depends heavily on uncertain parameters such as main sequence and post-main sequence mass loss, rotation and potentially unresolved physical phenomena \cite[e.g.][]{2021ApJ...923...41L}; and different sets of stellar evolution models predict different values for the threshold mass above which stripped-envelope stars can be formed from single-star evolution, particularly as a function of metallicity \citep[for a detailed discussion, see][]{2020A&A...634A..79S}.

\subsubsection{Physical and numerical uncertainties}

Uncertainties in the evolution of massive stars are known to appear in the late stages of evolution due to numerical resolution and the choice of nuclear network. We model energy generation rates and chemical abundances with MESA's \texttt{approx21} nuclear network, which is the smallest nuclear network that can be used to model massive stars to core collapse. This choice may affect the very late phases, but the lifetimes, compositions, and other global evolutionary properties discussed in this paper will not be affected \citep{2016ApJS..227...22F}.

The efficiency of convection and the presence of overshooting in helium stars are also uncertain. Varying these parameters may lead to more massive convective helium-burning cores. This will result in different lifetimes, and lead to significant differences in evolution, particularly in the lower mass regime. Overshooting in low mass helium stars will most significantly affect the locations of the boundaries in initial mass that lead to different final outcomes \citep{2022arXiv220100871C}, but since low mass helium stars (with luminosities below the WN limit) likely have very low mass loss rates regardless of their metallicity, we argue that the minimum mass at which stripped-envelope stars experience core collapse instead of becoming thermonuclear or electron-capture SNe is not strongly dependent on metallicity, and make a conservative choice for this limit, setting it at 3$\mso$.

Another uncertainty is introduced in our results from the implementation of MESA's \texttt{mlt++} scheme \citep{MESAII}. It artificially increases the efficiency of convective energy transport in the envelopes of stars, and is commonly implemented in calculations of massive star evolution to increase numerical stability. However, this tool also suppresses the effect of envelope inflation, which can occur in WR stars, as they are close to the Eddington limit \citep[e.g.][]{2015A&A...580A..20S,2006A&A...450..219P}. We argue that, since the mass-luminosity relation for WR stars is well defined \citep{1989A&A...210...93L}, and the empirically obtained mass loss rates do not depend on any other parameters defined at the surface of these stars, using \texttt{mlt++} will not have an impact on the core evolution of helium stars, or in the amount of mass that they lose due to winds. However, this will result in computed radii in our models that are likely underestimated. Furthermore, the proximity to the Eddington limit has been associated with violent mass eruptions, as in the case of luminous blue variable stars \citep[e.g.][]{2012A&A...538A..40G}. Therefore we argue that the final masses of our models, particularly those above the minimum WN luminosity, are an upper limit, and that the real distribution of SN ejecta masses is likely to be shifted to lower masses. Additionally, the ejecta masses of stripped-envelope SNe can be further displaced to lower values if helium stars expand significantly due to inflation, and experience a second episode of interaction with their companion \citep{2013ApJ...778L..23T}, or if they experience mass eruptions due to their high luminosities.

\subsubsection{Rotation in stripped-envelope stars}

The models presented in this paper are computed without accounting for the effects of rotation. Rotation can induce mixing of chemical elements in the stellar interior, thereby modifying its structure, evolution, surface composition and final outcome \citep[see, e.g.][]{2000ApJ...528..368H}. However, \cite{2010ApJ...725..940Y} found that angular momentum loss during mass transfer is significant, and most helium stars produced in binary systems by mass transfer are slow rotators. Helium stars produced in single stars, stripped by winds alone, lose most of their angular momentum in the winds, and therefore are also expected to be slow rotators \citep[e.g.][]{2003A&A...404..975M}. Therefore, we argue that the evolution of stripped-envelope stars is well approximated by non-rotating models of helium stars.

Some helium stars have been suggested to end their lives with significant rotation, particularly when stripping leads to short-period ($< 2$ days) binaries where tidal spin-up is particularly efficient\citep{2018A&A...616A..28Q}. However, this mainly affects low-metallicity systems, and the binary configuration that leads to their formation is rare.

Very fast rotation during early hydrogen burning can lead to chemically homogeneously evolving (CHE) stars, an evolutionary channel that diverges from canonical massive star evolution \citep{1987A&A...178..159M}. CHE stars will transport most of the hydrogen in their surfaces into their cores, where it is burnt into helium, and form critically rotating helium stars. CHE stars might be formed in mass accretors in binary systems at very low metallicity, but it is not expected to be common in the metallicity range considered in our models \citep{2007A&A...465L..29C}. They may also form in very close binaries, \citep{2009A&A...497..243D}, but are also expected to be rare and more prominent at metallicities lower than those of our models.

\cite{2019ApJ...878...49W} argues that helium stars approximately evolve like CHE stars. However, \cite{2018ApJ...858..115A,2020ApJ...901..114A} find that rotational mixing is still efficient after core hydrogen depletion. This allows the convective helium-burning core to grow, instead of decreasing in mass as in the case of helium stars (see Fig. \ref{fig:kips}), and prevents the formation of strong chemical composition gradients in the interior of CHE stars during and after core helium burning. This results in different structures and surface chemical compositions between helium stars and CHE stars throughout their evolution. Since they are more prevalent in low metallicity environments, CHE stars lose very little mass due to winds. However, they lose large fractions of mass (both steadily and in bursts) due the combined effect of contraction and rotation. This mass likely remains nearby the star, as it is lost via centrifugal acceleration, and likely contributes to their photometric and spectral properties. Therefore, we argue that their observable properties will also be significantly different. Furthermore, their core structures differ from those of non-rotating helium stars, and they lead to a different distribution of parameters such as core compactness and carbon abundance as a function of initial mass, that have an effect on their final outcome.

\subsubsection{The minimum luminosity of Wolf-Rayet stars}

The formalism presented in this work to calculate the optical depth at the surface of WR stars, adapted from \cite{1989A&A...210...93L}, makes several simplifying assumptions on the structure and opacity of WR winds. The choices of the values of parameters in the model, such as setting $\kappa$ to the electron scattering opacity, the density distribution of wind following a $\beta$ law with $\beta=1$, and the surface and terminal velocities are a simplification, since these may not represent WR stars accurately. For instance, the opacity at the base of the wind includes lines and may be a factor of few larger than the electron scattering opacity. Furthermore, its value is likely metallicity dependent. This contribution would also increase at higher metallicity. However, we justify the use of these choices with the agreement they produced with the observed values of the minimum luminosities of  WN type stars in the Galactic, LMC and SMC populations.

The mass-radius relation employed in this calculation \citep{1989A&A...210...93L} is calibrated for models of WR stars that do not experience inflation. Inflation can occur in luminous WR stars, or in WR stars with high envelope opacities \citep[e.g.][]{2006A&A...450..219P,2012A&A...538A..40G}. However, it becomes important only for evolved WR stars, and WR stars with luminosities that are considerably larger than the minimum luminosity WR stars predicted by our optical depth model, even at the highest metallicities in our grid, at least during the helium burning lifetime. The mass luminosity relation employed \citep{1989A&A...210...93L} was obtained through fits of hydrostatic, helium burning models. It was found to depend on composition, such that pure helium stars of a given luminosity were predicted to be less massive if they were enriched in carbon and oxygen. In our calculations, we employed only the mass-luminosity relation obtained for pure helium stars, but we found that employing a mass-luminosity relation for carbon and oxygen rich stars (Eq. 19 in \citealt{1989A&A...210...93L}) only changes the fit parameters of Eq. \ref{eq:fit_wc} by about 2\%.

The minimum luminosities of WN and WC stars are not trivial to determine from the observed samples of WR stars. WR stars are rare objects, since massive stars only spend a short fraction of their total lifetime in this phase. The populations in the Magellanic Clouds are thought to be complete, and there are only 12 WR stars in the SMC \citep{2003PASP..115.1265M}, and 154 in the LMC \citep{2018ApJ...863..181N}, compared with the hundreds of massive stars in the SMC \citep[e.g.][]{2018ApJ...868...57C} and thousands in the LMC \citep[e.g.][]{2013A&A...558A.134D}. The value of the minimum luminosity of WC stars is more complicated to determine because WC stars are much more scarce than WN stars, particularly at low metallicity. Out of the 12 WR stars in the SMC, 11 of them are WN type, and only 7 of them are hydrogen-poor (with $X_H$<0.3, but not completely hydrogen free). The value of the minimum luminosity of WC stars in the SMC is very uncertain since there are no WC stars in the SMC, and only one WO star. However, according to our estimate, this star appears to be close to the minimum luminosity limit. In the LMC, 28 of the 154 WR type stars are of type WC and WO, making this limit easier to determine.

More WR stars are found in the Galaxy \citep[e.g.][]{2019A&A...625A..57H,2019A&A...621A..92S}, but the population is still very incomplete, and parameters of many WR stars --such as mass loss rate, surface composition, temperature and bolometric luminosity-- have not been determined due to the lack of high resolution spectra. Furthermore, there is a significant metallicity gradient in the Galaxy \citep{2017A&A...600A..70A}, which increases the uncertainty in determining the values of the minimum luminosities of WN and WC stars within the Galaxy.

Despite the uncertainties that go into the determination of the minimum luminosities of WN and WC stars, both theoretically and observationally, we argue that our results highlight the importance of taking the luminosity distribution of WR stars in different metallicities into account, as well as the mass loss rates that these stars experience during the WN and WC evolutionary stages. The fits provided in Eqs. \ref{eq:fit_wn} and \ref{eq:fit_wc} can be employed in stellar evolution models and population synthesis studies that aim to study WR stars to properly account for WR mass loss rates, and to distinguish between helium stars with optically thin winds and WR stars of different types.

\subsubsection{Uncertainties in our stellar population models}

The calculations obtained from the population model described in Sect. \ref{sec:methods_popsynth} are a crude approximation, since not every star that is formed becomes stripped, and this probability likely depends on ZAMS mass. Furthermore, many WRs will likely have evolved from single stars that perhaps retain a hydrogen envelope during a substantial fraction of their helium burning lifetime, and WRs that evolve from binaries are neither always stripped exactly at the beginning of helium burning, nor necessarily completely stripped. According to \cite{pauli}, WR stars in the LMC are most commonly formed by binary stripping. Therefore, our models are a good proxy for most of the formed WR stars, at least at this metallicity. Additional stripped-envelope stars are likely formed by reverse-stripping, when the least massive component of a binary system, which is not accounted for by \cite{pauli}. The number of systems that undergo this process, however, is uncertain, as it depends on natal kicks \citep{2020PASA...37...38V}. The approximations made in our population model can help us guide our understanding of how varying strengths of WR winds can influence the components of a stellar population.

Since we assume that stars are fully stripped by the beginning of core helium burning, the fractional lifetime that stars spend as a WC star represents an upper limit (unless other effects such as a second mass transfer occur, as suggested by \cite{2020A&A...637A...6L} for sub-solar metallicity). However, this general trend occurs regardless of the extent of remaining hydrogen shells, because even if the WR lifetime is only a fraction of the helium burning lifetime, the effect of metallicity will be the same \citep{2005A&A...429..581M}.

As discussed in Sect. \ref{sec:populations}, in the metallicity range that our simulations cover, every time a star has its carbon and oxygen rich layers exposed, it will also be luminous enough to appear as a WC star. Therefore, the ratio of WC to WN stars is set by the metallicity dependence of the minimum luminosity of WN stars, the efficiency of WN winds in forming WC stars, the lifetimes of WR stars in each phase, and the stripping probability of stars as a function of mass. In our population model, we assume that this efficiency is constant as a function of mass, effectively isolating the influence of winds and the minimum WN luminosity in shaping a population of WR stars.

\subsection{Comparison to previous stellar evolution models}

The study of the late phases of evolution of stripped-envelope stars is becoming a very active field of stellar astrophysics. Stripped-envelope stars have been modelled as single helium stars \citep{1995ApJ...448..315W,2017MNRAS.470.3970Y,2018MNRAS.481.1908K,2019ApJ...878...49W,2021MNRAS.505.4874H}, and as the late phases of evolution of massive single stars \citep{2002ApJ...580L..35M,2005A&A...429..581M}, although models of full binary evolution are quickly becoming available as well \citep{2017PASA...34...58E,2021A&A...645A...5S,2021A&A...656A..58L}. The results of these simulations have evolved significantly in recent years, partly because our understanding of WR mass loss has increased considerably in recent years.

We include the WR mass loss rates of \cite{2017MNRAS.470.3970Y}, which differ from what most works on stripped-envelope stars have employed, particularly in that we consider that winds of WN and WC stars are different. However, in agreement with \cite{2019ApJ...878...49W,2021A&A...645A...5S} and \cite{2021A&A...656A..58L}, we find that the retreat in convective helium core mass has a large effect on the evolution of stripped-envelope stars.

The final masses of our models with $Z=0.02$ are in agreement with \cite{2017MNRAS.470.3970Y} and \cite{2020A&A...642A.106D}. The final masses and lifetimes of our $Z=0.015$ models are also consistent with the $Z=0.0145$ models from \cite{2019ApJ...878...49W} at masses where the WC/WO wind is unimportant. However, we observe systematically lower final masses for helium stars that evolve as WC/WO than what is found by \cite{2019ApJ...878...49W}, at similar metallicity. The core evolution, carbon-oxygen core masses and luminosities of our models agrees well with the results of \cite{2019ApJ...878...49W} in every mass regime, despite the fact that he employed a much more complete nuclear network in his calculations.

The theoretically computed mass loss rates of stripped-envelope stars of \cite{2020MNRAS.491.4406S} naturally account for the transition between WR stars with strong mass loss rates and transparent wind helium stars, which they predicted to have mass loss rates that are several orders of magnitude lower than those of WR stars. They find that the transition between them is not abrupt, but the mass loss rates of stripped-envelope stars are predicted to quickly decrease below a certain luminosity. They also find that the metallicity dependence of mass loss rates is less steep than the one employed in this work. Therefore, the helium star models of \cite{2021MNRAS.505.4874H}, calculated using the mass loss rates of \cite{2020MNRAS.491.4406S}, yield higher final masses than those of \cite{2017MNRAS.470.3970Y}, \cite{2019ApJ...878...49W}, and the models in this work. However, this becomes significant only in the range of masses of the most massive BH progenitors, and the discrepancies in the mass range where the grids overlap are small.

Although the mass loss rates of WR stars employed by \cite{2021MNRAS.505.4874H} are different from those used in this work, the consensus is that erroneous results can be obtained in the analysis of stripped-envelope stars due to potential caveats, which result from extrapolating the empirically derived mass loss rates of WR stars beyond their applicability range, as is the case for progenitors of gravitational wave sources.

A detailed analysis of the core properties of our models, and a comparison with other works that have addressed this in the context of progenitors of Type I SNe is deferred to \citetalias{aguilera-dena2021}. Some of the ramifications for compact-object mergers are also discussed in \cite{2022A&A...657L...6A}.

\section{Conclusions}\label{sec:conclusions}

We have modelled grids of evolutionary sequences of helium stars to mimic the late evolution of massive stars that have been stripped of their hydrogen envelopes. Envelope stripping in massive stars can occur if they are in binary or multiple systems, where a companion is close enough to have an interaction (including stable mass transfer and CE ejections). Alternatively, stars can lose their envelopes via stellar winds or massive ejections during and after the main sequence, but this is only effective for stars with the highest initial masses, and those born in high metallicity environments. We have covered the range of initial helium masses between 1.5 and 70 $\mso$, and metallicities between 0.01, similar to that observed in the LMC, to 0.04, similar to the environments of highest metallicity known, such as the region near the centre of our Galaxy, or in nearby massive galaxies.

Evolved helium stars contain a strong chemical composition gradient. Layers initially outside the convective helium burning core 
contain abundant nitrogen, enhanced during hydrogen burning due to CNO processing. The convective helium-burning core, on the other hand, is nitrogen-free and has a progressively decreasing amount of helium, and will instead be enriched in carbon and oxygen. This gives rise to the observational dichotomy between WN and WC stars. 
We follow \citep{2015A&A...581A.110T} in assuming that
WO stars are formed from stripped-envelope stars that have finished core helium burning, and expose oxygen-rich layers.

Through our models, we find the evolution of massive stars is deeply affected by stripping, and by the amount of mass loss they experience after stripping, which is larger at high metallicity.
We characterised the lifetime of helium stars, finding that it is weakly dependent on metallicity. However, the duration of the phases where our models spend time as WN, WC and WO type candidates is strongly dependent on metallicity, and will determine the composition of WR populations across the Universe. We found that the minimum initial helium star mass at which stars develop a WC phase, and a WO phase, varies strongly with metallicity.

We find that the duration of the WO phase is mediated by the remaining lifetime after core helium depletion, and the time at which oxygen rich layers become exposed, both of which increase with metallicity. Therefore, the minimum initial helium star mass at which models are found to spend some time in the WO phase also increases with increasing metallicity.

We propose a method to find the minimum luminosity of WR stars, which is different for WN and WC type stars, as a function of metallicity. This is accomplished by combining the empirical mass loss rates of \cite{2017MNRAS.470.3970Y} and the wind optical depth, as derived by \cite{1989A&A...210...93L}. We provide a quantitative way to determine this limit in different environments. 

Using this method, we find that many stars in our grid will not experience a WN type phase, but rather evolve as transparent wind helium stars. We also find that helium stars that are enriched with carbon and oxygen in their surface will always be observable as WR stars (i.e., will have optically thick winds) at the metallicities that we cover in our models. However, extrapolating 
to low metallicities, we find that stars with transparent winds and with surface layers composed of helium burning products, should exist \citep[akin to the models of][]{2015A&A...581A..15S}.

Our models reproduce the WC star luminosity distributions in the LMC and our Galaxy well, except for very few low-luminosity outliers. Even those have luminosities considerably above the predicted transition between optically thin and optically thick winds. Possibly, these stars not only a product of isolated helium star evolution, but could be formed through rare binary evolution channels, e.g. white dwarf merger \citep{2019Natur.569..684G},
or instabilities in the envelopes of some helium stars.

Through simplified population synthesis, we find that the number of stripped-envelope stars increases slightly with metallicity. We find that transparent-wind stripped-envelope stars outnumber WR stars by a considerable fraction. Using the metallicity in our grids that most closely resembles the metallicity of the LMC, we predict that it will contain about 350 transparent-wind stripped-envelope stars, which may eventually produce Type Ib SNe. Furthermore, the expected number ratio of WC to WN type stars as a function of metallicity, and find that it to increase as a function of metallicity, since stars at higher metallicities spend a longer fraction of their life as WC stars, even after taking into account that the number of WN stars might increase due to the decrease of the WN star luminosity limit. However, this increase is not dramatic and stalls at $Z\sim0.03$ due to the shorter lifetime of WC stars, compared to WN stars.

We also compute the expected number ratio of WO to WN stars, and of WO to WC stars. Both quantities increase with increasing metallicity, because the lifetimes of WO stars increase as a function of metallicity, while the minimum mass up to which they are formed decreases.
Our estimates are in agreement with the observed number of WO stars in the LMC. For M31 and M33, our models predict some 2 to 3 WO type stars,
while none have been found so far. 
We emphasise that these number ratios are sensitive to the minimum WN star luminosity, which has not been considered in previous population synthesis works.

With this study, we find that populations of stripped-envelope stars will not only be determined by the efficiency of their formation channels, i.e., main-sequence star winds and binary interactions. They will also be influenced by the metallicity dependent minimum luminosity at which stripped-envelope stars develop optically thick winds and WR characteristics. 
Furthermore, the metallicity dependent wind mass loss of stripped-envelope 
stars is key in determining the WR sub-type distribution. In combination with our findings in \citetalias{aguilera-dena2021} and \cite{2022A&A...657L...6A}, we emphasise the need for a holistic approach to the study of the evolution of stripped-envelope stars, both through observations of individual systems and stellar populations, and through the study of stripped-envelope SNe across cosmic time.

\begin{acknowledgements}
D. R. A.-D. and J. A. are supported by the Stavros Niarchos Foundation (SNF) and the Hellenic Foundation for Research and Innovation (H.F.R.I.) under the 2nd Call of ``Science and Society’' Action Always strive for excellence – ``Theodoros Papazoglou’' (Project Number: 01431). DP acknowledges financial support by the Deutsches Zentrum für Luft und Raumfahrt (DLR) grant FKZ 50 OR 2005. SCY has been supported by the National Research Foundation of Korea (NRF)
grant (NRF-2019R1A2C2010885).

D. R. A.-D. would like to acknowledge valuable discussions with Eva Laplace and Jakub Klencki.

This research made extensive use of NASA's ADS and \textsc{mesa}\footnote{\url{http://mesastar.org}} \citep{MESAI,MESAII,MESAIII,2018ApJS..234...34P}.
\end{acknowledgements}

\bibliographystyle{aa}
\bibliography{references}

\begin{thebibliography}{123}
\expandafter\ifx\csname natexlab\endcsname\relax\def\natexlab#1{#1}\fi

\bibitem[{{Aguilera-Dena} {et~al.}(2020){Aguilera-Dena}, {Langer},
  {Antoniadis}, \& {M{\"u}ller}}]{2020ApJ...901..114A}
{Aguilera-Dena}, D.~R., {Langer}, N., {Antoniadis}, J., \& {M{\"u}ller}, B.
  2020, \apj, 901, 114

\bibitem[{{Aguilera-Dena} {et~al.}(2022){Aguilera-Dena}, Langer, Antoniadis,
  M{\"u}ller, Dessart, Vigna-Gomez, \& Yoon}]{aguilera-dena2021}
{Aguilera-Dena}, D.~R., Langer, N., Antoniadis, J., {et~al.} 2022, Astronomy \&
  Astrophysics, to be submitted

\bibitem[{{Aguilera-Dena} {et~al.}(2018){Aguilera-Dena}, {Langer}, {Moriya}, \&
  {Schootemeijer}}]{2018ApJ...858..115A}
{Aguilera-Dena}, D.~R., {Langer}, N., {Moriya}, T.~J., \& {Schootemeijer}, A.
  2018, \apj, 858, 115

\bibitem[{{Anders} {et~al.}(2017){Anders}, {Chiappini}, {Minchev}, {Miglio},
  {Montalb{\'a}n}, {Mosser}, {Rodrigues}, {Santiago}, {Baudin}, {Beers}, {da
  Costa}, {Garc{\'\i}a}, {Garc{\'\i}a-Hern{\'a}ndez}, {Holtzman}, {Maia},
  {Majewski}, {Mathur}, {Noels-Grotsch}, {Pan}, {Schneider}, {Schultheis},
  {Steinmetz}, {Valentini}, \& {Zamora}}]{2017A&A...600A..70A}
{Anders}, F., {Chiappini}, C., {Minchev}, I., {et~al.} 2017, \aap, 600, A70

\bibitem[{{Antoniadis} {et~al.}(2022){Antoniadis}, {Aguilera-Dena},
  {Vigna-G{\'o}mez}, {Kramer}, {Langer}, {M{\"u}ller}, {Tauris}, {Wang}, \&
  {Xu}}]{2022A&A...657L...6A}
{Antoniadis}, J., {Aguilera-Dena}, D.~R., {Vigna-G{\'o}mez}, A., {et~al.} 2022,
  \aap, 657, L6

\bibitem[{{Antoniadis} {et~al.}(2020){Antoniadis}, {Chanlaridis},
  {Gr{\"a}fener}, \& {Langer}}]{2020A&A...635A..72A}
{Antoniadis}, J., {Chanlaridis}, S., {Gr{\"a}fener}, G., \& {Langer}, N. 2020,
  \aap, 635, A72

\bibitem[{{Arnett}(1974)}]{1974ApJ...194..373A}
{Arnett}, W.~D. 1974, \apj, 194, 373

\bibitem[{{Asplund} {et~al.}(2009){Asplund}, {Grevesse}, {Sauval}, \&
  {Scott}}]{2009ARA&A..47..481A}
{Asplund}, M., {Grevesse}, N., {Sauval}, A.~J., \& {Scott}, P. 2009, \araa, 47,
  481

\bibitem[{{Bartzakos} {et~al.}(2001){Bartzakos}, {Moffat}, \&
  {Niemela}}]{2001MNRAS.324...18B}
{Bartzakos}, P., {Moffat}, A.~F.~J., \& {Niemela}, V.~S. 2001, \mnras, 324, 18

\bibitem[{{B{\"o}hm-Vitense}(1958)}]{1958ZA.....46..108B}
{B{\"o}hm-Vitense}, E. 1958, \zap, 46, 108

\bibitem[{{Brown} {et~al.}(2001){Brown}, {Heger}, {Langer}, {Lee}, {Wellstein},
  \& {Bethe}}]{2001NewA....6..457B}
{Brown}, G.~E., {Heger}, A., {Langer}, N., {et~al.} 2001, \na, 6, 457

\bibitem[{{Cantiello} \& {Langer}(2010)}]{2010A&A...521A...9C}
{Cantiello}, M. \& {Langer}, N. 2010, \aap, 521, A9

\bibitem[{{Cantiello} {et~al.}(2007){Cantiello}, {Yoon}, {Langer}, \&
  {Livio}}]{2007A&A...465L..29C}
{Cantiello}, M., {Yoon}, S.~C., {Langer}, N., \& {Livio}, M. 2007, \aap, 465,
  L29

\bibitem[{{Castro} {et~al.}(2018){Castro}, {Oey}, {Fossati}, \&
  {Langer}}]{2018ApJ...868...57C}
{Castro}, N., {Oey}, M.~S., {Fossati}, L., \& {Langer}, N. 2018, \apj, 868, 57

\bibitem[{{Chanlaridis} {et~al.}(2022){Chanlaridis}, {Antoniadis},
  {Aguilera-Dena}, {Gr{\"a}fener}, {Langer}, \&
  {Stergioulas}}]{2022arXiv220100871C}
{Chanlaridis}, S., {Antoniadis}, J., {Aguilera-Dena}, D.~R., {et~al.} 2022,
  arXiv e-prints, arXiv:2201.00871

\bibitem[{{Conti}(1975)}]{1975MSRSL...9..193C}
{Conti}, P.~S. 1975, Memoires of the Societe Royale des Sciences de Liege, 9,
  193

\bibitem[{{Crowther}(2007)}]{2007ARA&A..45..177C}
{Crowther}, P.~A. 2007, \araa, 45, 177

\bibitem[{{de Mink} {et~al.}(2009){de Mink}, {Cantiello}, {Langer}, {Pols},
  {Brott}, \& {Yoon}}]{2009A&A...497..243D}
{de Mink}, S.~E., {Cantiello}, M., {Langer}, N., {et~al.} 2009, \aap, 497, 243

\bibitem[{{Dessart} {et~al.}(2020){Dessart}, {Yoon}, {Aguilera-Dena}, \&
  {Langer}}]{2020A&A...642A.106D}
{Dessart}, L., {Yoon}, S.-C., {Aguilera-Dena}, D.~R., \& {Langer}, N. 2020,
  \aap, 642, A106

\bibitem[{{Doran} {et~al.}(2013){Doran}, {Crowther}, {de Koter}, {Evans},
  {McEvoy}, {Walborn}, {Bastian}, {Bestenlehner}, {Gr{\"a}fener}, {Herrero},
  {K{\"o}hler}, {Ma{\'\i}z Apell{\'a}niz}, {Najarro}, {Puls}, {Sana},
  {Schneider}, {Taylor}, {van Loon}, \& {Vink}}]{2013A&A...558A.134D}
{Doran}, E.~I., {Crowther}, P.~A., {de Koter}, A., {et~al.} 2013, \aap, 558,
  A134

\bibitem[{{Eldridge} {et~al.}(2017){Eldridge}, {Stanway}, {Xiao}, {McClelland},
  {Taylor}, {Ng}, {Greis}, \& {Bray}}]{2017PASA...34...58E}
{Eldridge}, J.~J., {Stanway}, E.~R., {Xiao}, L., {et~al.} 2017, \pasa, 34, e058

\bibitem[{{Ertl} {et~al.}(2020){Ertl}, {Woosley}, {Sukhbold}, \&
  {Janka}}]{2020ApJ...890...51E}
{Ertl}, T., {Woosley}, S.~E., {Sukhbold}, T., \& {Janka}, H.~T. 2020, \apj,
  890, 51

\bibitem[{{Evans} {et~al.}(2011){Evans}, {Taylor}, {H{\'e}nault-Brunet},
  {Sana}, {de Koter}, {Sim{\'o}n-D{\'\i}az}, {Carraro}, {Bagnoli}, {Bastian},
  {Bestenlehner}, {Bonanos}, {Bressert}, {Brott}, {Campbell}, {Cantiello},
  {Clark}, {Costa}, {Crowther}, {de Mink}, {Doran}, {Dufton}, {Dunstall},
  {Friedrich}, {Garcia}, {Gieles}, {Gr{\"a}fener}, {Herrero}, {Howarth},
  {Izzard}, {Langer}, {Lennon}, {Ma{\'\i}z Apell{\'a}niz}, {Markova},
  {Najarro}, {Puls}, {Ramirez}, {Sab{\'\i}n-Sanjuli{\'a}n}, {Smartt}, {Stroud},
  {van Loon}, {Vink}, \& {Walborn}}]{2011A&A...530A.108E}
{Evans}, C.~J., {Taylor}, W.~D., {H{\'e}nault-Brunet}, V., {et~al.} 2011, \aap,
  530, A108

\bibitem[{{Farmer} {et~al.}(2016){Farmer}, {Fields}, {Petermann}, {Dessart},
  {Cantiello}, {Paxton}, \& {Timmes}}]{2016ApJS..227...22F}
{Farmer}, R., {Fields}, C.~E., {Petermann}, I., {et~al.} 2016, \apjs, 227, 22

\bibitem[{{Farmer} {et~al.}(2020){Farmer}, {Renzo}, {de Mink}, {Fishbach}, \&
  {Justham}}]{2020ApJ...902L..36F}
{Farmer}, R., {Renzo}, M., {de Mink}, S.~E., {Fishbach}, M., \& {Justham}, S.
  2020, \apjl, 902, L36

\bibitem[{{Fragos} {et~al.}(2019){Fragos}, {Andrews}, {Ramirez-Ruiz}, {Meynet},
  {Kalogera}, {Taam}, \& {Zezas}}]{2019ApJ...883L..45F}
{Fragos}, T., {Andrews}, J.~J., {Ramirez-Ruiz}, E., {et~al.} 2019, \apjl, 883,
  L45

\bibitem[{{Georgy} {et~al.}(2012){Georgy}, {Ekstr{\"o}m}, {Meynet}, {Massey},
  {Levesque}, {Hirschi}, {Eggenberger}, \& {Maeder}}]{2012A&A...542A..29G}
{Georgy}, C., {Ekstr{\"o}m}, S., {Meynet}, G., {et~al.} 2012, \aap, 542, A29

\bibitem[{{Gilkis} {et~al.}(2019){Gilkis}, {Vink}, {Eldridge}, \&
  {Tout}}]{2019MNRAS.486.4451G}
{Gilkis}, A., {Vink}, J.~S., {Eldridge}, J.~J., \& {Tout}, C.~A. 2019, \mnras,
  486, 4451

\bibitem[{{G{\"o}tberg} {et~al.}(2018){G{\"o}tberg}, {de Mink}, {Groh},
  {Kupfer}, {Crowther}, {Zapartas}, \& {Renzo}}]{2018A&A...615A..78G}
{G{\"o}tberg}, Y., {de Mink}, S.~E., {Groh}, J.~H., {et~al.} 2018, \aap, 615,
  A78

\bibitem[{{G{\"o}tberg} {et~al.}(2020){G{\"o}tberg}, {de Mink}, {McQuinn},
  {Zapartas}, {Groh}, \& {Norman}}]{2020A&A...634A.134G}
{G{\"o}tberg}, Y., {de Mink}, S.~E., {McQuinn}, M., {et~al.} 2020, \aap, 634,
  A134

\bibitem[{{Gr{\"a}fener} \& {Hamann}(2005)}]{2005A&A...432..633G}
{Gr{\"a}fener}, G. \& {Hamann}, W.~R. 2005, \aap, 432, 633

\bibitem[{{Gr{\"a}fener} \& {Hamann}(2008)}]{2008A&A...482..945G}
{Gr{\"a}fener}, G. \& {Hamann}, W.~R. 2008, \aap, 482, 945

\bibitem[{{Gr{\"a}fener} {et~al.}(2017){Gr{\"a}fener}, {Owocki}, {Grassitelli},
  \& {Langer}}]{2017A&A...608A..34G}
{Gr{\"a}fener}, G., {Owocki}, S.~P., {Grassitelli}, L., \& {Langer}, N. 2017,
  \aap, 608, A34

\bibitem[{{Gr{\"a}fener} {et~al.}(2012){Gr{\"a}fener}, {Owocki}, \&
  {Vink}}]{2012A&A...538A..40G}
{Gr{\"a}fener}, G., {Owocki}, S.~P., \& {Vink}, J.~S. 2012, \aap, 538, A40

\bibitem[{{Gr{\"a}fener} {et~al.}(2011){Gr{\"a}fener}, {Vink}, {de Koter}, \&
  {Langer}}]{2011A&A...535A..56G}
{Gr{\"a}fener}, G., {Vink}, J.~S., {de Koter}, A., \& {Langer}, N. 2011, \aap,
  535, A56

\bibitem[{{Grevesse} {et~al.}(1996){Grevesse}, {Noels}, \&
  {Sauval}}]{1996ASPC...99..117G}
{Grevesse}, N., {Noels}, A., \& {Sauval}, A.~J. 1996, in Astronomical Society
  of the Pacific Conference Series, Vol.~99, Cosmic Abundances, ed. S.~S.
  {Holt} \& G.~{Sonneborn}, 117

\bibitem[{{Groh} {et~al.}(2014){Groh}, {Meynet}, {Ekstr{\"o}m}, \&
  {Georgy}}]{2014A&A...564A..30G}
{Groh}, J.~H., {Meynet}, G., {Ekstr{\"o}m}, S., \& {Georgy}, C. 2014, \aap,
  564, A30

\bibitem[{{Gvaramadze} {et~al.}(2019){Gvaramadze}, {Gr{\"a}fener}, {Langer},
  {Maryeva}, {Kniazev}, {Moskvitin}, \& {Spiridonova}}]{2019Natur.569..684G}
{Gvaramadze}, V.~V., {Gr{\"a}fener}, G., {Langer}, N., {et~al.} 2019, \nat,
  569, 684

\bibitem[{{Hainich} {et~al.}(2015){Hainich}, {Pasemann}, {Todt}, {Shenar},
  {Sander}, \& {Hamann}}]{2015A&A...581A..21H}
{Hainich}, R., {Pasemann}, D., {Todt}, H., {et~al.} 2015, \aap, 581, A21

\bibitem[{{Hainich} {et~al.}(2014){Hainich}, {R{\"u}hling}, {Todt}, {Oskinova},
  {Liermann}, {Gr{\"a}fener}, {Foellmi}, {Schnurr}, \&
  {Hamann}}]{2014A&A...565A..27H}
{Hainich}, R., {R{\"u}hling}, U., {Todt}, H., {et~al.} 2014, \aap, 565, A27

\bibitem[{{Hamann} {et~al.}(2006){Hamann}, {Gr{\"a}fener}, \&
  {Liermann}}]{2006A&A...457.1015H}
{Hamann}, W.~R., {Gr{\"a}fener}, G., \& {Liermann}, A. 2006, \aap, 457, 1015

\bibitem[{{Hamann} {et~al.}(2019){Hamann}, {Gr{\"a}fener}, {Liermann},
  {Hainich}, {Sander}, {Shenar}, {Ramachandran}, {Todt}, \&
  {Oskinova}}]{2019A&A...625A..57H}
{Hamann}, W.~R., {Gr{\"a}fener}, G., {Liermann}, A., {et~al.} 2019, \aap, 625,
  A57

\bibitem[{{Heger} {et~al.}(2000){Heger}, {Langer}, \&
  {Woosley}}]{2000ApJ...528..368H}
{Heger}, A., {Langer}, N., \& {Woosley}, S.~E. 2000, \apj, 528, 368

\bibitem[{{Higgins} {et~al.}(2021){Higgins}, {Sander}, {Vink}, \&
  {Hirschi}}]{2021MNRAS.505.4874H}
{Higgins}, E.~R., {Sander}, A.~A.~C., {Vink}, J.~S., \& {Hirschi}, R. 2021,
  \mnras, 505, 4874

\bibitem[{{Hillier} {et~al.}(2021){Hillier}, {Aadland}, {Massey}, \&
  {Morrell}}]{2021MNRAS.503.2726H}
{Hillier}, D.~J., {Aadland}, E., {Massey}, P., \& {Morrell}, N. 2021, \mnras,
  503, 2726

\bibitem[{{Ivanova}(2011)}]{2011ApJ...730...76I}
{Ivanova}, N. 2011, \apj, 730, 76

\bibitem[{{Klencki} {et~al.}(2021){Klencki}, {Istrate}, {Nelemans}, \&
  {Pols}}]{2021arXiv211110271K}
{Klencki}, J., {Istrate}, A.~G., {Nelemans}, G., \& {Pols}, O. 2021, arXiv
  e-prints, arXiv:2111.10271

\bibitem[{{Klencki} {et~al.}(2020){Klencki}, {Nelemans}, {Istrate}, \&
  {Pols}}]{2020A&A...638A..55K}
{Klencki}, J., {Nelemans}, G., {Istrate}, A.~G., \& {Pols}, O. 2020, \aap, 638,
  A55

\bibitem[{{Kruckow} {et~al.}(2018){Kruckow}, {Tauris}, {Langer}, {Kramer}, \&
  {Izzard}}]{2018MNRAS.481.1908K}
{Kruckow}, M.~U., {Tauris}, T.~M., {Langer}, N., {Kramer}, M., \& {Izzard},
  R.~G. 2018, \mnras, 481, 1908

\bibitem[{{Kruckow} {et~al.}(2016){Kruckow}, {Tauris}, {Langer}, {Sz{\'e}csi},
  {Marchant}, \& {Podsiadlowski}}]{2016A&A...596A..58K}
{Kruckow}, M.~U., {Tauris}, T.~M., {Langer}, N., {et~al.} 2016, \aap, 596, A58

\bibitem[{{Langer}(1989)}]{1989A&A...210...93L}
{Langer}, N. 1989, \aap, 210, 93

\bibitem[{{Langer}(1991)}]{1991A&A...248..531L}
{Langer}, N. 1991, \aap, 248, 531

\bibitem[{{Langer}(2012)}]{2012ARA&A..50..107L}
{Langer}, N. 2012, \araa, 50, 107

\bibitem[{{Langer} {et~al.}(1994){Langer}, {Hamann}, {Lennon}, {Najarro},
  {Pauldrach}, \& {Puls}}]{1994A&A...290..819L}
{Langer}, N., {Hamann}, W.~R., {Lennon}, M., {et~al.} 1994, \aap, 290, 819

\bibitem[{{Laplace} {et~al.}(2020){Laplace}, {G{\"o}tberg}, {de Mink},
  {Justham}, \& {Farmer}}]{2020A&A...637A...6L}
{Laplace}, E., {G{\"o}tberg}, Y., {de Mink}, S.~E., {Justham}, S., \& {Farmer},
  R. 2020, \aap, 637, A6

\bibitem[{{Laplace} {et~al.}(2021){Laplace}, {Justham}, {Renzo}, {G{\"o}tberg},
  {Farmer}, {Vartanyan}, \& {de Mink}}]{2021A&A...656A..58L}
{Laplace}, E., {Justham}, S., {Renzo}, M., {et~al.} 2021, \aap, 656, A58

\bibitem[{{Law-Smith} {et~al.}(2020){Law-Smith}, {Everson}, {Ramirez-Ruiz}, {de
  Mink}, {van Son}, {G{\"o}tberg}, {Zellmann}, {Vigna-G{\'o}mez}, {Renzo},
  {Wu}, {Schr{\o}der}, {Foley}, \& {Hutchinson-Smith}}]{2020arXiv201106630L}
{Law-Smith}, J. A.~P., {Everson}, R.~W., {Ramirez-Ruiz}, E., {et~al.} 2020,
  arXiv e-prints, arXiv:2011.06630

\bibitem[{{Leung} {et~al.}(2021){Leung}, {Wu}, \&
  {Fuller}}]{2021ApJ...923...41L}
{Leung}, S.-C., {Wu}, S., \& {Fuller}, J. 2021, \apj, 923, 41

\bibitem[{{Maeder}(1981)}]{1981A&A...102..401M}
{Maeder}, A. 1981, \aap, 102, 401

\bibitem[{{Maeder}(1987)}]{1987A&A...178..159M}
{Maeder}, A. 1987, \aap, 178, 159

\bibitem[{{Massey} \& {Holmes}(2002)}]{2002ApJ...580L..35M}
{Massey}, P. \& {Holmes}, S. 2002, \apjl, 580, L35

\bibitem[{{Massey} \& {Johnson}(1998)}]{1998ApJ...505..793M}
{Massey}, P. \& {Johnson}, O. 1998, \apj, 505, 793

\bibitem[{{Massey} {et~al.}(2014){Massey}, {Neugent}, {Morrell}, \&
  {Hillier}}]{2014ApJ...788...83M}
{Massey}, P., {Neugent}, K.~F., {Morrell}, N., \& {Hillier}, D.~J. 2014, \apj,
  788, 83

\bibitem[{{Massey} {et~al.}(2003){Massey}, {Olsen}, \&
  {Parker}}]{2003PASP..115.1265M}
{Massey}, P., {Olsen}, K.~A.~G., \& {Parker}, J.~W. 2003, \pasp, 115, 1265

\bibitem[{{McClelland} \& {Eldridge}(2016)}]{2016MNRAS.459.1505M}
{McClelland}, L.~A.~S. \& {Eldridge}, J.~J. 2016, \mnras, 459, 1505

\bibitem[{{Meynet} \& {Maeder}(2003)}]{2003A&A...404..975M}
{Meynet}, G. \& {Maeder}, A. 2003, \aap, 404, 975

\bibitem[{{Meynet} \& {Maeder}(2005)}]{2005A&A...429..581M}
{Meynet}, G. \& {Maeder}, A. 2005, \aap, 429, 581

\bibitem[{{Modjaz} {et~al.}(2019){Modjaz}, {Guti{\'e}rrez}, \&
  {Arcavi}}]{2019NatAs...3..717M}
{Modjaz}, M., {Guti{\'e}rrez}, C.~P., \& {Arcavi}, I. 2019, Nature Astronomy,
  3, 717

\bibitem[{{Moe} \& {Di Stefano}(2017)}]{2017ApJS..230...15M}
{Moe}, M. \& {Di Stefano}, R. 2017, \apjs, 230, 15

\bibitem[{{Moffat}(1989)}]{1989ApJ...347..373M}
{Moffat}, A. F.~J. 1989, \apj, 347, 373

\bibitem[{{Neugent} \& {Massey}(2019)}]{2019Galax...7...74N}
{Neugent}, K. \& {Massey}, P. 2019, Galaxies, 7, 74

\bibitem[{{Neugent} \& {Massey}(2011)}]{2011ApJ...733..123N}
{Neugent}, K.~F. \& {Massey}, P. 2011, \apj, 733, 123

\bibitem[{{Neugent} {et~al.}(2012){Neugent}, {Massey}, \&
  {Georgy}}]{2012ApJ...759...11N}
{Neugent}, K.~F., {Massey}, P., \& {Georgy}, C. 2012, \apj, 759, 11

\bibitem[{{Neugent} {et~al.}(2018){Neugent}, {Massey}, \&
  {Morrell}}]{2018ApJ...863..181N}
{Neugent}, K.~F., {Massey}, P., \& {Morrell}, N. 2018, \apj, 863, 181

\bibitem[{{Nomoto} \& {Hashimoto}(1988)}]{1988PhR...163...13N}
{Nomoto}, K. \& {Hashimoto}, M. 1988, \physrep, 163, 13

\bibitem[{{Pantaleoni Gonz{\'a}lez} {et~al.}(2021){Pantaleoni Gonz{\'a}lez},
  {Ma{\'\i}z Apell{\'a}niz}, {Barb{\'a}}, \& {Reed}}]{2021MNRAS.504.2968P}
{Pantaleoni Gonz{\'a}lez}, M., {Ma{\'\i}z Apell{\'a}niz}, J., {Barb{\'a}},
  R.~H., \& {Reed}, B.~C. 2021, \mnras, 504, 2968

\bibitem[{{Pauli} {et~al.}(2022){Pauli}, Langer, Marchant, Aguilera-Dena, \&
  Wang}]{pauli}
{Pauli}, D., Langer, N., Marchant, P., Aguilera-Dena, D.~R., \& Wang, C. 2022,
  Astronomy \& Astrophysics, to be submitted

\bibitem[{{Paxton} {et~al.}(2011){Paxton}, {Bildsten}, {Dotter}, {Herwig},
  {Lesaffre}, \& {Timmes}}]{MESAI}
{Paxton}, B., {Bildsten}, L., {Dotter}, A., {et~al.} 2011, \apjs, 192, 3

\bibitem[{{Paxton} {et~al.}(2013){Paxton}, {Cantiello}, {Arras}, {Bildsten},
  {Brown}, {Dotter}, {Mankovich}, {Montgomery}, {Stello}, {Timmes}, \&
  {Townsend}}]{MESAII}
{Paxton}, B., {Cantiello}, M., {Arras}, P., {et~al.} 2013, \apjs, 208, 4

\bibitem[{{Paxton} {et~al.}(2015){Paxton}, {Marchant}, {Schwab}, {Bauer},
  {Bildsten}, {Cantiello}, {Dessart}, {Farmer}, {Hu}, {Langer}, {Townsend},
  {Townsley}, \& {Timmes}}]{MESAIII}
{Paxton}, B., {Marchant}, P., {Schwab}, J., {et~al.} 2015, \apjs, 220, 15

\bibitem[{{Paxton} {et~al.}(2018){Paxton}, {Schwab}, {Bauer}, {Bildsten},
  {Blinnikov}, {Duffell}, {Farmer}, {Goldberg}, {Marchant}, {Sorokina},
  {Thoul}, {Townsend}, \& {Timmes}}]{2018ApJS..234...34P}
{Paxton}, B., {Schwab}, J., {Bauer}, E.~B., {et~al.} 2018, \apjs, 234, 34

\bibitem[{{Peimbert} {et~al.}(2007){Peimbert}, {Luridiana}, {Peimbert}, \&
  {Carigi}}]{2007ASPC..374...81P}
{Peimbert}, M., {Luridiana}, V., {Peimbert}, A., \& {Carigi}, L. 2007, in
  Astronomical Society of the Pacific Conference Series, Vol. 374, From Stars
  to Galaxies: Building the Pieces to Build Up the Universe, ed.
  A.~{Vallenari}, R.~{Tantalo}, L.~{Portinari}, \& A.~{Moretti}, 81

\bibitem[{{Petrovic} {et~al.}(2006){Petrovic}, {Pols}, \&
  {Langer}}]{2006A&A...450..219P}
{Petrovic}, J., {Pols}, O., \& {Langer}, N. 2006, \aap, 450, 219

\bibitem[{{Podsiadlowski} {et~al.}(1992){Podsiadlowski}, {Joss}, \&
  {Hsu}}]{1992ApJ...391..246P}
{Podsiadlowski}, P., {Joss}, P.~C., \& {Hsu}, J.~J.~L. 1992, \apj, 391, 246

\bibitem[{{Qin} {et~al.}(2018){Qin}, {Fragos}, {Meynet}, {Andrews},
  {S{\o}rensen}, \& {Song}}]{2018A&A...616A..28Q}
{Qin}, Y., {Fragos}, T., {Meynet}, G., {et~al.} 2018, \aap, 616, A28

\bibitem[{{Salpeter}(1955)}]{1955ApJ...121..161S}
{Salpeter}, E.~E. 1955, \apj, 121, 161

\bibitem[{{Sana} {et~al.}(2012){Sana}, {de Mink}, {de Koter}, {Langer},
  {Evans}, {Gieles}, {Gosset}, {Izzard}, {Le Bouquin}, \&
  {Schneider}}]{2012Sci...337..444S}
{Sana}, H., {de Mink}, S.~E., {de Koter}, A., {et~al.} 2012, Science, 337, 444

\bibitem[{{Sana} \& {Evans}(2011)}]{2011IAUS..272..474S}
{Sana}, H. \& {Evans}, C.~J. 2011, in IAU Symposium, Vol. 272, Active OB Stars:
  Structure, Evolution, Mass Loss, and Critical Limits, ed. C.~{Neiner},
  G.~{Wade}, G.~{Meynet}, \& G.~{Peters}, 474--485

\bibitem[{{Sana} {et~al.}(2014){Sana}, {Le Bouquin}, {Lacour}, {Berger},
  {Duvert}, {Gauchet}, {Norris}, {Olofsson}, {Pickel}, {Zins}, {Absil}, {de
  Koter}, {Kratter}, {Schnurr}, \& {Zinnecker}}]{2014ApJS..215...15S}
{Sana}, H., {Le Bouquin}, J.~B., {Lacour}, S., {et~al.} 2014, \apjs, 215, 15

\bibitem[{{Sander} {et~al.}(2019){Sander}, {Hamann}, {Todt}, {Hainich},
  {Shenar}, {Ramachandran}, \& {Oskinova}}]{2019A&A...621A..92S}
{Sander}, A.~A.~C., {Hamann}, W.~R., {Todt}, H., {et~al.} 2019, \aap, 621, A92

\bibitem[{{Sander} \& {Vink}(2020)}]{2020MNRAS.499..873S}
{Sander}, A. A.~C. \& {Vink}, J.~S. 2020, \mnras, 499, 873

\bibitem[{{Sander} {et~al.}(2020){Sander}, {Vink}, \&
  {Hamann}}]{2020MNRAS.491.4406S}
{Sander}, A. A.~C., {Vink}, J.~S., \& {Hamann}, W.~R. 2020, \mnras, 491, 4406

\bibitem[{{Sanyal} {et~al.}(2015){Sanyal}, {Grassitelli}, {Langer}, \&
  {Bestenlehner}}]{2015A&A...580A..20S}
{Sanyal}, D., {Grassitelli}, L., {Langer}, N., \& {Bestenlehner}, J.~M. 2015,
  \aap, 580, A20

\bibitem[{{Schmutz} {et~al.}(1989){Schmutz}, {Hamann}, \&
  {Wessolowski}}]{1989A&A...210..236S}
{Schmutz}, W., {Hamann}, W.~R., \& {Wessolowski}, U. 1989, \aap, 210, 236

\bibitem[{{Schneider} {et~al.}(2021){Schneider}, {Podsiadlowski}, \&
  {M{\"u}ller}}]{2021A&A...645A...5S}
{Schneider}, F.~R.~N., {Podsiadlowski}, P., \& {M{\"u}ller}, B. 2021, \aap,
  645, A5

\bibitem[{{Schootemeijer} {et~al.}(2019){Schootemeijer}, {Langer}, {Grin}, \&
  {Wang}}]{2019A&A...625A.132S}
{Schootemeijer}, A., {Langer}, N., {Grin}, N.~J., \& {Wang}, C. 2019, \aap,
  625, A132

\bibitem[{{Sen} {et~al.}(2021){Sen}, {Langer}, {Marchant}, {Menon}, {de Mink},
  {Schootemeijer}, {Sch{\"u}rmann}, {Mahy}, {Hastings}, {Nathaniel}, {Sana},
  {Wang}, \& {Xu}}]{2021arXiv211103329S}
{Sen}, K., {Langer}, N., {Marchant}, P., {et~al.} 2021, arXiv e-prints,
  arXiv:2111.03329

\bibitem[{{Shenar} {et~al.}(2020){Shenar}, {Gilkis}, {Vink}, {Sana}, \& {Sand
  er}}]{2020A&A...634A..79S}
{Shenar}, T., {Gilkis}, A., {Vink}, J.~S., {Sana}, H., \& {Sand er}, A.~A.~C.
  2020, \aap, 634, A79

\bibitem[{{Shenar} {et~al.}(2016){Shenar}, {Hainich}, {Todt}, {Sander},
  {Hamann}, {Moffat}, {Eldridge}, {Pablo}, {Oskinova}, \&
  {Richardson}}]{2016A&A...591A..22S}
{Shenar}, T., {Hainich}, R., {Todt}, H., {et~al.} 2016, \aap, 591, A22

\bibitem[{{Smith} {et~al.}(1994){Smith}, {Meynet}, \&
  {Mermilliod}}]{1994A&A...287..835S}
{Smith}, L.~F., {Meynet}, G., \& {Mermilliod}, J.~C. 1994, \aap, 287, 835

\bibitem[{{Sukhbold} \& {Woosley}(2014)}]{2014ApJ...783...10S}
{Sukhbold}, T. \& {Woosley}, S.~E. 2014, \apj, 783, 10

\bibitem[{{Sz{\'e}csi} {et~al.}(2015){Sz{\'e}csi}, {Langer}, {Yoon}, {Sanyal},
  {de Mink}, {Evans}, \& {Dermine}}]{2015A&A...581A..15S}
{Sz{\'e}csi}, D., {Langer}, N., {Yoon}, S.-C., {et~al.} 2015, \aap, 581, A15

\bibitem[{{Tauris} \& {Dewi}(2001)}]{2001A&A...369..170T}
{Tauris}, T.~M. \& {Dewi}, J.~D.~M. 2001, \aap, 369, 170

\bibitem[{{Tauris} {et~al.}(2013){Tauris}, {Langer}, {Moriya}, {Podsiadlowski},
  {Yoon}, \& {Blinnikov}}]{2013ApJ...778L..23T}
{Tauris}, T.~M., {Langer}, N., {Moriya}, T.~J., {et~al.} 2013, \apjl, 778, L23

\bibitem[{{Tauris} {et~al.}(2015){Tauris}, {Langer}, \&
  {Podsiadlowski}}]{2015MNRAS.451.2123T}
{Tauris}, T.~M., {Langer}, N., \& {Podsiadlowski}, P. 2015, \mnras, 451, 2123

\bibitem[{{Tauris} \& {van den Heuvel}(2006)}]{2006csxs.book..623T}
{Tauris}, T.~M. \& {van den Heuvel}, E.~P.~J. 2006, {Formation and evolution of
  compact stellar X-ray sources}, Vol.~39, 623--665

\bibitem[{{Tramper} {et~al.}(2016){Tramper}, {Sana}, \& {de
  Koter}}]{2016ApJ...833..133T}
{Tramper}, F., {Sana}, H., \& {de Koter}, A. 2016, \apj, 833, 133

\bibitem[{{Tramper} {et~al.}(2015){Tramper}, {Straal}, {Sanyal}, {Sana}, {de
  Koter}, {Gr{\"a}fener}, {Langer}, {Vink}, {de Mink}, \&
  {Kaper}}]{2015A&A...581A.110T}
{Tramper}, F., {Straal}, S.~M., {Sanyal}, D., {et~al.} 2015, \aap, 581, A110

\bibitem[{{Vanbeveren} {et~al.}(2007){Vanbeveren}, {Van Bever}, \&
  {Belkus}}]{2007ApJ...662L.107V}
{Vanbeveren}, D., {Van Bever}, J., \& {Belkus}, H. 2007, \apjl, 662, L107

\bibitem[{{Vigna-G{\'o}mez} {et~al.}(2020){Vigna-G{\'o}mez}, {MacLeod},
  {Neijssel}, {Broekgaarden}, {Justham}, {Howitt}, {de Mink}, {Vinciguerra}, \&
  {Mandel}}]{2020PASA...37...38V}
{Vigna-G{\'o}mez}, A., {MacLeod}, M., {Neijssel}, C.~J., {et~al.} 2020, \pasa,
  37, e038

\bibitem[{{Vigna-G{\'o}mez} {et~al.}(2018){Vigna-G{\'o}mez}, {Neijssel},
  {Stevenson}, {Barrett}, {Belczynski}, {Justham}, {de Mink}, {M{\"u}ller},
  {Podsiadlowski}, {Renzo}, {Sz{\'e}csi}, \& {Mandel}}]{2018MNRAS.481.4009V}
{Vigna-G{\'o}mez}, A., {Neijssel}, C.~J., {Stevenson}, S., {et~al.} 2018,
  \mnras, 481, 4009

\bibitem[{{Vigna-G{\'o}mez} {et~al.}(2021){Vigna-G{\'o}mez}, {Wassink},
  {Klencki}, {Istrate}, {Nelemans}, \& {Mandel}}]{2021arXiv210714526V}
{Vigna-G{\'o}mez}, A., {Wassink}, M., {Klencki}, J., {et~al.} 2021, arXiv
  e-prints, arXiv:2107.14526

\bibitem[{{Vink}(2017)}]{2017A&A...607L...8V}
{Vink}, J.~S. 2017, \aap, 607, L8

\bibitem[{{Vink} \& {de Koter}(2005)}]{2005A&A...442..587V}
{Vink}, J.~S. \& {de Koter}, A. 2005, \aap, 442, 587

\bibitem[{{Webbink}(1984)}]{1984ApJ...277..355W}
{Webbink}, R.~F. 1984, \apj, 277, 355

\bibitem[{{Wellstein} {et~al.}(2001){Wellstein}, {Langer}, \&
  {Braun}}]{2001A&A...369..939W}
{Wellstein}, S., {Langer}, N., \& {Braun}, H. 2001, \aap, 369, 939

\bibitem[{{Woosley}(2017)}]{2017ApJ...836..244W}
{Woosley}, S.~E. 2017, \apj, 836, 244

\bibitem[{{Woosley}(2019)}]{2019ApJ...878...49W}
{Woosley}, S.~E. 2019, \apj, 878, 49

\bibitem[{{Woosley} {et~al.}(1995){Woosley}, {Langer}, \&
  {Weaver}}]{1995ApJ...448..315W}
{Woosley}, S.~E., {Langer}, N., \& {Weaver}, T.~A. 1995, \apj, 448, 315

\bibitem[{{Woosley} {et~al.}(2020){Woosley}, {Sukhbold}, \&
  {Janka}}]{2020ApJ...896...56W}
{Woosley}, S.~E., {Sukhbold}, T., \& {Janka}, H.~T. 2020, \apj, 896, 56

\bibitem[{{Yoon}(2017)}]{2017MNRAS.470.3970Y}
{Yoon}, S.-C. 2017, \mnras, 470, 3970

\bibitem[{{Yoon} {et~al.}(2017){Yoon}, {Dessart}, \&
  {Clocchiatti}}]{2017ApJ...840...10Y}
{Yoon}, S.-C., {Dessart}, L., \& {Clocchiatti}, A. 2017, \apj, 840, 10

\bibitem[{{Yoon} {et~al.}(2010){Yoon}, {Woosley}, \&
  {Langer}}]{2010ApJ...725..940Y}
{Yoon}, S.-C., {Woosley}, S.~E., \& {Langer}, N. 2010, \apj, 725, 940

\end{thebibliography}

\end{document}